\documentclass[preprint,review, 12pt]{elsarticle}

\usepackage{graphicx}

\usepackage{amssymb}
\usepackage{amsbsy}
\usepackage[version=3]{mhchem}

\usepackage[nodots]{numcompress}

\newcommand{\au}{\,\mathrm{a.u.}}
\newcommand{\vecs}[1]{\boldsymbol{#1}} 

\journal{Chemical Physics}

\begin{document}

\begin{frontmatter}


\title{Optimization Schemes for Selective Molecular Cleavage with Tailored Ultrashort Laser Pulses}



\author[1]{Kevin Krieger}
\author[2]{Alberto Castro\corref{cor1}}
\ead{acastro@bifi.es}
\author[1]{E. K. U. Gross}

\cortext[cor1]{Corresponding author}

\address[1]{Max-Planck-Institut f{\"{u}}r
    Mikrostrukturphysik, Weinberg 2, D-06120 Halle, Germany.}
\address[2]{Institute for Biocomputation and Physics
    of Complex Systems (BIFI), University of Zaragoza, E-50018
    Zaragoza, Spain.}

\begin{abstract}
We present some approaches to the computation of ultra-fast laser
pulses capable of selectively breaking molecular bonds. The
calculations are based on a mixed quantum-classical description: The
electrons are treated quantum mechanically (making use of
time-dependent density-functional theory), whereas the nuclei are
treated classically.  The temporal shape of the pulses is tailored to
maximise a control target functional which is designed to produce the
desired molecular cleavage. The precise definition of this functional
is a crucial ingredient: we explore expressions based on the forces,
on the momenta and on the velocities of the nuclei. The algorithm used
to find the optimum pulse is also relevant; we test both direct
gradient-free algorithms, as well as schemes based on formal optimal
control theory. The tests are performed both on one dimensional models
of atomic chains, and on first-principles descriptions of molecules.
\end{abstract}


\end{frontmatter}


\section{Introduction}
\label{intro}

Soon after its first operation,\cite{MAIMAN1960} the laser was
expected to become the ultimate surgical tool at the nanoscopic level:
Light, at convenient wave-lengths, monochromatic, coherent, and
intense,\cite{CHANG2005} was believed to open the avenue to
selectively break (or create) molecular bonds. Unfortunately, the
early attempts to perform this kind of photo-chemistry were only
occasionally successful.\cite{doi:10.1021/j150667a004,zewail:27} These
attempts used ``simple'' monochromatic lasers, tinkering only with two
parameters: the frequency and the intensity. However, the energy,
tuned to a particular vibrational frequency and initially deposited on
the corresponding bond, is soon re-distributed to the rest of the
modes, and produces undesired global heating instead of selective
cleavage.\cite{doi:10.1146/annurev.pc.42.100191.000503}

The ``controlled'' laser assisted photo-chemistry advanced along with
improvements on laser technology, with methodologies such as the
control of quantum interference proposed by Brumer and
Shapiro,\cite{Brumer1986541,Brumer1986177,shapiro-2003} the
``pump-dump'' control proposed by Tannor and
Rice,\cite{Tannor1985541,Herek199415} stimulated Raman adiabatic
passage,\cite{Gaubatz1988463} wave-packet
interferometry,\cite{doi:10.1146/annurev.physchem.59.032607.093818}
and others.\cite{1367-2630-11-10-105030, Brif2010} The key ingredients,
beyond mere mono-chromaticity and intensity, were shown to be
coherence (and therefore, interference), detailed shaping, and
ultra-short pulse duration (in the femto-second time scale). The most
successful technique is adaptive feedback control (AFC), as proposed
by Judson and Rabitz,\cite{PhysRevLett.68.1500} and first realised in
1997.\cite{Bardeen1997151}

There are two important components in an AFC experiment: the pulse
shaper,\cite{Weiner2000} and the search algorithm fed by the repeated
measurement outcome. The former is an instrument that allows to almost
arbitrarily design laser pulses. The increasing versatility of modern
laser sources (regarding pulse length, power, and accessible
frequencies), and the capacity of pulse shapers to modify the produced
pulses, set the boundaries that theoretical studies such as the one
presented in this work must respect; however these boundaries are
rapidly pushed further, allowing more versatile pulses.

Quantum optimal control theory
(QOCT)\cite{doi:10.1088/0953-4075/40/18/R01, shi19886870,
  peirce19884950, kosloff1989201} is the most general theoretical
framework aimed to the prediction of laser pulses that are optimal for
a given task. It is the translation to the quantum realm of a very
broad mathematical area, optimal control, that is best formulated in
the language of systems theory.\cite{luenberger1979,luenberger1969}
Its use for quantum processes was initiated in the
80s\cite{shi19886870, peirce19884950, kosloff1989201} -- responding to
the initial experimental stir. In some way, QOCT encompasses all the
previously mentioned optimisation methods (inasmuch as it may describe
them theoretically). The theory is constructed on top of some chosen
level of approximation for the description of the process that is to
be optimised. Here lies the main limitation of QOCT:\cite{Brif2010} it
may only be predictive if the system is simple enough to allow for an
accurate approximation of its evolution. In most cases, however, the
process is too complex.

If some reliable predictive power is to be expected from any QOCT
calculation, one should attempt a first-principles description. In
particular, in the regime of interest, the dynamics of the electrons
should be carefully treated: high intensity electric fields at high
frequencies affect directly the electronic degrees of freedom.
Indeed, when many-electron systems are irradiated with strong
femtosecond pulses a number of interesting non-trivial photo-reactions
may take place: above-threshold or tunnel ionisation, bond hardening or
softening, high harmonic generation, photo-isomerisation,
photo-fragmentation, Coulomb explosion, etc.\cite{protopapas1997389,
  brabec2000545, scrinzi2006R1}. Yet most of the computational work
until now has relied on simplified models, and has usually worked with
nuclear wave packets -- defined on a few relevant reaction variables,
after a reduction of dimensionality has been postulated -- moving on
one or a few Born-Oppenheimer potential energy surfaces, and therefore
mostly ignoring the dynamic behaviour of the electrons. Direct,
first-principles, electronic control has been scarcely
attempted,\cite{doi:797291} unless for one-electron
cases.\cite{doi:10.1209/0295-5075/87/53001,
  doi:10.1103/PhysRevLett.98.157404, doi:10.1103/PhysRevB.77.085324}

One viable alternative to treat electronic motion in an ab initio way
is time-dependent density-functional theory
(TDDFT).\cite{Marques2006-book, PhysRevLett.52.997} Recently some of
us have demonstrated the feasibility of performing QOCT with
TDDFT.\cite{arxiv:1009.2241v1} This was not obvious due to the
non-linear character of the TDDFT equations: the usual QOCT equations
assume a standard, linear Schr{\"{o}}dinger-like evolution, and the
resulting QOCT equations are correspondingly simple. However, the
presence of the Hartree, exchange and correlation term in the TDDFT
equations need special care.

TDDFT offers reasonable accuracy when dealing with the non-linear
response of molecular systems, with a fraction of the cost of methods
based on the wave function. Furthermore, the electronic system
described within TDDFT may then be coupled to the ionic motion in a
mixed quantum-classical description.\cite{saalmann1996153,
  saalmann19983213, gross1996, doi:10.1140/epjd/e2003-00306-3} This
model will obviously ignore quantum nuclear effects, but may be
sufficient for the description of many processes. In this work we
present our first results based on this combination. In Section
\ref{section:methodology}, the essential equations are displayed, as
well as a brief description of the numerical procedure.  Section
\ref{section:forces} describes the results of the optimisations when
the target functional is defined in terms of the values of the forces
on the nuclei at the end of the laser pulse, for 1D models, whereas in
Section \ref{section:momenta}, the target functional is defined in
terms of the momenta. In Section \ref{section:chains}, the attempt to
selectively break molecular chains is described. Finally, Sections
\ref{section:h3} and \ref{section:formaldimine} display results for
fully ab initio 3D calculations.

\section{Methodology}
\label{section:methodology}

\subsection{Essentials of QOCT}

We consider a quantum mechanical system governed by
Schr{\"{o}}dinger's equation during the time interval [0,T] (atomic
units will be used hereafter):
\begin{eqnarray}
\label{eq:schroedinger-1}
i\frac{\partial \Psi}{\partial t} (x,t)  & = & \hat{H}[u,t] \Psi(x,t) \,,
\\
\label{eq:schroedinger-2}
\Psi(x,0) & = & \Psi_0(x)\,,
\end{eqnarray}
where $x$ is the full set of quantum coordinates, and $u$ is a
\emph{control}, typically a set of parameters that determine the
precise shape of an external potential applied to the
system. Mathematically, we can distinguish two types of
``representation'' for the control $u$:
\begin{enumerate}
\item $u$ is a real valued continuous function defined on the time
  interval of interest (the \emph{control function}); we will call this
  a ``real-time'' representation of the control. For example, the Hamiltonian
  may have the form:
\begin{equation}
\label{eq:hamiltonian-ufunction}
  \hat{H}[u,t] = \hat{H}_0 + u(t)\hat{D}\,.
\end{equation}
\item $u$ is a set of $N$ real parameters that modifies the precise
  shape of the Hamiltonian; typically, this set of parameters fixes
  the form of a control function; we will call this a
  ``parameterised'' representation of the control.
\end{enumerate}
In any case, the specification of $u$, together with an initial value
condition, $\Psi(0)=\Psi_0$ determines the full evolution of the system,
$\Psi[u]$, via the propagation of Schr{\"{o}}dinger's equation.

We wish to maximize the function $G$,
\begin{equation}
G[u] = F[\Psi[u],u]\,,
\end{equation}
where $F$ is the so-called ``target functional''; in many cases it is
split into two parts, $F[\Psi,u] = J_1[\Psi] + J_2[u]$, so that $J_1$
only depends on the state of the system, and $J_2$ is called the
``penalty'', and depends explicitly on the control $u$. An important
distinction should be made regarding $J_1$:
\begin{enumerate}
\item It may depend on the full evolution of the system during the
  time interval [0,T]; this is usually called a time-dependent
  target. We may write this as $J_1[\Psi] = J_1^{[0,T]}[\Psi]$, where
  the $J_1^{[0,T]}[\Psi]$ functional admits continuous functional
  derivatives, in particular $\displaystyle \frac{\delta
    J_1^{[0,T]}}{\delta \Psi^*(x,t)}$ is continuous at $t=T$.
\item $J_1$ may only depend on the state of the system at the end of
  the propagation, which we may write as $J_1[\Psi] = J_1^T[\Psi(T)]$.
\end{enumerate}
Of course, $J_1$ may be defined as a combination of the two options, i.e.:
\begin{equation}
J_1[\Psi] = J_1^{[0,T]}[\Psi] + J_1^T[\Psi(T)]\,.
\end{equation}
Note that, in this case:
\begin{equation}
\frac{\delta J_1}{\delta \Psi^*(x,t)} = \frac{\delta J_1^{[0,T]}}{\delta \Psi^*(x,t)} + \delta(t-T)
\frac{\delta J_1^T}{\delta \Psi^*(x,T)}\,.
\end{equation}

In most cases these functionals are defined as the expectation value
of some observable $\hat{O}$. For example:
\begin{equation}
J_1^{T}[\Psi(T)] = \langle \Psi(T)\vert\hat{O}\vert\Psi(T)\rangle\,,\;{\rm or:}
\end{equation}
\begin{equation}
J_1^{[0,T]}[\Psi(T)] = \int_0^T\!\!\!{\rm d}t\; \langle \Psi(t)\vert\hat{O}(t)\vert\Psi(t)\rangle\,.
\end{equation}

One needs now an optimization algorithm to find the maximum (or
maxima) of $G$. Two broad families can be distinguished: gradient-free
procedures, that only require some means to compute the value of $G$
given a control input $u$, and gradient-based procedures, that also
necessitate the computation of the gradient of $G$ with respect to $u$
(more precisely, the functional derivative if $u$ is a continuous
function in time). We will not repeat here a derivation that can be
found elsewhere in several forms;\cite{peirce19884950, kosloff1989201,
  doi:10.1088/0953-4075/40/18/R01, luenberger1969,
  doi:10.1063/1.1650297, doi:10.1103/PhysRevA.71.053810} the key
equations are:
\begin{eqnarray}
\nonumber
\nabla_u G[u] & = & \left.\nabla_u F[\Psi,u]\right|_{\Psi = \Psi[u]} + 
\\\label{eq:qoct-gradient}
& & 2 {\rm Im} \int_0^T\!\!\!{\rm d}t\; \langle \chi[u](t) \vert \nabla_u \hat{H}[u,t] \vert \Psi[u](t)\rangle\,,
\end{eqnarray}
in case $u$ is a set of real parameters, and:
\begin{eqnarray}
\nonumber
\frac{\delta G}{\delta u(t)} & = & \left. \frac{\delta F[\Psi,u]}{\delta u(t)}\right|_{\Psi = \Psi[u]} + 
\\\label{eq:qoct-functional-derivative}
& & 2 {\rm Im} \langle \chi[u](t) \vert \hat{D} \vert \Psi[u](t)\rangle\,,
\end{eqnarray}
if $u$ is a function in time, and the Hamiltonian is given by Eq.~\ref{eq:hamiltonian-ufunction}.

Note that a new ``wave function'', $\chi[u]$, has been introduced;
it is given by the solution of:
\begin{eqnarray}
\label{eq:lambda-1}
i\frac{\partial \chi[u]}{\partial t} (x,t)  & = & \hat{H}^{\dagger}[u,t] \chi[u](x,t) 
-i \frac{\delta J_1^{[0,T]}}{\delta \Psi^*[u](x,t)}\,,
\\
\label{eq:lambda-2}
\chi[u](x,T) & = & \frac{\delta J_1^T}{\delta \Psi^*[u](x,T)} \,.
\end{eqnarray}
This is similar to the original Schr{\"{o}}dinger's equation
(Eqs.~\ref{eq:schroedinger-1} and ~\ref{eq:schroedinger-2}), except:
(1) It may be inhomogeneous, if $J_1^{[0,T]}$ is not zero (i.e. if the
target is time-dependent~\cite{doi:10.1063/1.1650297,
  doi:10.1103/PhysRevA.71.053810}), and (2) The initial condition is
given at the final time $t=T$, which implies it must be propagated
\emph{backwards}.

The computation of the gradient or functional derivative of G,
therefore, requires $\Psi[u]$ and $\chi[u]$, which are obtained by
first propagating Eq.~\ref{eq:schroedinger-1} forwards, and then
Eq.~\ref{eq:lambda-1} backwards. The maxima of $G$ are found at the
critical points $\nabla_u G[u] = 0$ or $\frac{\delta G}{\delta u(t)} =
0$; in order to arrive to these maxima one can use a variety of
algorithms, some of which are listed in Section~\ref{sec:methodology-numerical}.

\subsection{Mixed quantum-classical description with TDDFT}

Instead of solving the many-electron Schr{\"{o}}dinger equation, TDDFT
allows to work with a set of one-electron equations, the Kohn-Sham
(KS) system, corresponding to a fictitious system whose one-particle
density is by construction identical to that of the real one:
\begin{eqnarray}
\nonumber
\i\frac{\partial \varphi_i}{\partial t}(\vec{r},t) & = & 
-\frac{1}{2}\nabla^2\varphi_i(\vec{r},t) + \left[ v_{\rm ext}(\vec{r},t) \right.
\\
& & + \left. v_{\rm Hartree}[n_t](\vec{r}) + v_{\rm xc}[n](\vec{r},t) \right] \varphi_i(\vec{r},t)\,,
\\
n(\vec{r},t) & = & \sum_{i=1}^N 2\vert\varphi_i(\vec{r},t)\vert^2 \equiv n_t(\vec{r}) \,.
\end{eqnarray}
We will assume a system with $2N$ electrons in a spin compensated
configuration, evolving in a spin independent Hamiltonian. This means
$N$ doubly occupied KS orbitals $\varphi_i$, $i=1,\dots,N$. The system
evolves on an external time-dependent potential $v_{\rm ext}$, that
may include the interaction with a set of nuclei, as well as external
electric fields. The Hartree term $v_{\rm Hartree}$ is the classic
electrostatic potential, and the rest of the electron-electron
interaction is encoded in the exchange and correlation potential
$v_{\rm xc}$. In this work, we will only use the adiabatic extension
of the local density approximation
(LDA),\cite{doi:10.1103/PhysRevLett.45.566} although the extension to
other more sophisticated schemes is straightforward.

The external potential can depend on a control function, and therefore
control theory can be employed to find optimal evolutions of the KS
system. Note, however, that the KS equations are not akin to the
conventional Schr{\"{o}}dinger equation, since they are
non-linear. The QOCT expressions derived above are therefore not
valid; the correct equations have been presented
elsewhere;\cite{arxiv:1009.2241v1} however, in this work we will
either (1) take the independent electron approximation, which amounts
to ignoring the mentioned non-linearity, for model calculations, or
(2) utilize a gradient-free version of QOCT, for which we can use the
full-fledged version of TDDFT.

In order to describe the combined coupled movement of electrons and
(classical) nuclei, one can perform Ehrenfest dynamics on top of
TDDFT.\cite{doi:10.1140/epjd/e2003-00306-3} The external term $v_{\rm
  ext}$ will couple the electrons to $N_{\rm nuc}$ nuclei located at
positions $\vec{R}_{\alpha}(t)$ through an expression in the form:
\begin{equation}
v_{\rm ext}(\vec{r},t;\lbrace \vec{R}_\beta(t)\rbrace) = 
\sum_{\alpha=1}^{N_{\rm nuc}} \frac{-z_\alpha}{\vert \vec{R}_\alpha(t)-\vec{r}\vert} + \vec{E}(t)\cdot\vec{r}
\end{equation}
The evolution of the nuclear positions is then governed by an Ehrenfest
equation in the form:
\begin{eqnarray}\nonumber
m_\alpha\frac{{\rm d}^2}{{\rm d}t^2}\vec{R}_\alpha(t) & = & \sum_{\beta=1}^{N_{\rm nuc}} z_\alpha z_\beta
\frac{\vec{R}_\alpha(t)-\vec{R}_\beta(t)}{\vert \vec{R}_\alpha(t)-\vec{R}_\beta(t) \vert^3} + z_\alpha \vec{E}(t)
\\
& & - \int\!\!{\rm d}^3r\;n(\vec{r},t)\nabla_{\vec{R}_\alpha} v_{\rm ext}(\vec{r},t;\lbrace \vec{R}_\beta(t)\rbrace)\,.
\end{eqnarray}

\subsection{Numerical implementation}

All the ideas described above have been implemented in the {\tt
  octopus} code. Since the numerical details of this platform are
described elsewhere,\cite{Marques200360, Castro2006} here we will only
list some essential points. The laser field and the optimization
algorithms are described below with more detail.

\begin{itemize}

\item Wave functions and densities are represented on a regular
  rectangular real space mesh. This is a suitable scheme to describe
  high intensity laser-electron interactions, since the electronic
  density visits regions in space far from the localized basis sets
  typically used in other schemes. Furthermore, the intrinsic locality
  allows for easy parallelisation, and the only parameters controlling
  convergence are the grid spacing and the simulation box size.

\item The electron-ion interaction is modelled with
  pseudopotentials. In this way, the Coulomb singularity is avoided,
  and the core electrons are removed from the calculation. For the
  first results described below, however, we will use 1D models, and
  the soft-Coulomb interaction to avoid singularities.

\item The KS orbitals are evolved in real time with the help of a
  number of propagating algorithms.\cite{doi:10.1063/1.1774980} This
  is crucial since all algorithms require multiple propagations.

\item The code performs realistic 3D calculations, but it also allows
  1D and 2D models, such as the ones we will present below.

\end{itemize}

\subsubsection{The laser field.} 
\label{section:laser-field}

We will assume that laser pulses can be described in the dipole
approximation, which is valid given the wave lengths and intensities
that will be considered. In consequence, it suffices with an electric
field in the form:
\begin{equation}
\vec{E}(t) = \epsilon(t)\vec{p}\,,
\end{equation}
where $\vec{p}$ is a unit vector that determines the polarization
direction, and $\epsilon(t)$ determines the temporal dependence, and
is the object to be optimized -- i.e. the \emph{control function}.

Not any function in time is admissible as a solution; there are
physical and experimental constraints that must be respected. For
example, an important physical constraint is:
\begin{equation}
\label{eq:constraint-zero}
\int_0^T\!\! {\rm d}t\;  \epsilon(t)=0. 
\end{equation}
This condition follows from Maxwell's equations for a freely
propagating pulse in the electric dipole
approximation.\cite{chelkowski2005053815} Also, the pulses must
obviously start and end at zero:
\begin{equation}
\label{eq:zerofield}
\epsilon(0) = \epsilon(T) = 0\,.
\end{equation}
It is important to reduce the search space to functions that are
experimentally accessible, which means a limitation on the accessible
frequency components, and on the intensities. Regarding the latter,
usually it is done by considering the integrated intensity or
\emph{fluence}, defined as:
\begin{equation}
\label{eq:fluence}
\mathcal{F}[\epsilon] = \int_{0}^T\!\!\!\! {\rm d}t\; \epsilon^2(t)\,.
\end{equation}
Spectral constraints can also be imposed either by penalizing the
undesired frequencies in the definition of the
target,\cite{doi:10.1088/1464-4266/7/10/014} or by restricting from
the start the search space to the correct subspace.

As discussed earlier, we may use a real-time representation, and
therefore $\epsilon(t)$ is directly the control object $u$, or a
parameterised representation, in which this control function
$\epsilon(t)$ is determined by a set of parameters $u$. This
distinction is relevant for the mathematical derivations (since in the
former case functional derivatives must be used, whereas in the latter
case one uses normal gradients). Numerically, however, a function in
real time must also be discretized, and therefore the distinction
disappears. Nevertheless, typically the number of degrees of freedom
(number of grid points in time) will be much larger, and therefore the
algorithms utilized will differ.

Regarding the choices for the parameterisation, it is a natural choice
to expand the control field in a basis set, and to establish the
coefficients of this expansion as the parameters:
\begin{equation}
\epsilon(t) = \sum_{n=1}^N \tilde{\epsilon}_n g_n(t).
\label{eq:parameter_description}
\end{equation}
$N$ is the dimension of the real basis set $\lbrace g_n(t)\rbrace$. It
is chosen to be orthonormal over the interval $[0,T]$:
\begin{equation}
\int_0^T\!\!\!\! {\rm d}t\; g_m(t)g_n(t) = \delta_{mn}\,.
\end{equation}
In our calculations, two basis sets have been used: a sine basis:
\begin{equation}
g_n(t)=\sqrt{\frac{2}{T}}\sin(\frac{\pi}{T}n t)\,, \qquad n=1\dots N\,,
\label{eq:sinebasis}
\end{equation}
or a normal Fourier basis:
\begin{equation}
g_n(t) = \Bigg\{
\begin{array}{lll}
\sqrt{\frac{2}{T}}\cos(\frac{2\pi}{T}n t)\,,
\quad n=1,\dots ,\frac{N}{2} \\
\sqrt{\frac{2}{T}}\sin(\frac{2\pi}{T}(n-\frac{N}{2}) t)\,,
\quad n=(\frac{N}{2}+1),\dots ,N .
\end{array}
\label{eq:fourier_basis}
\end{equation}
The representation in these basis sets has the advantage that spectral
constraints can be automatically enforced: the maximum frequency is
given by the choice of $N$, and we we will not include the
zero-frequency component, in order to satisfy condition
(\ref{eq:constraint-zero}).

We can directly choose the basis set expansion coefficients as
constrol parameters, or else constrain further the search space to
meet other physical or experimental requirements, by defining the
coefficients as functions of a reduced set of parameters:
$\tilde{\epsilon}_n = \tilde{\epsilon}_n[u]$. Our choices have been
the following:

\begin{itemize} 

\item A \emph{constrained sine series}. The sine series,
  Eq.~\ref{eq:sinebasis} automatically fulfills the condition given by
  Eq.~(\ref{eq:zerofield}). To meet condition
  (\ref{eq:constraint-zero}), however, the following relation would
  have to be fulfilled:
\begin{equation}
\sum_{m=0}^{N/2-1} \frac{\tilde{\epsilon}_{(2m+1)}}{(2m+1)} = 0. \label{eq:sine_zero_const}
\end{equation}
For some of the cases presented below, we also enforced a fixed fluence. 
As function of any orthonormal basis set coefficients, the fluence is given by:
\begin{equation}
\mathcal{F}[\tilde{\epsilon}] = \sum_{n=1}^{N}\tilde{\epsilon}_n^2 . \label{eq:hyper_const}
\end{equation}
Setting the fluence to a predefined value $\mathcal{F}_0$ amounts to
requiring the vector $\tilde{\epsilon}$ to belong to a hypersphere. We
may then transform $\tilde{\epsilon}$ into hyperspherical coordinates;
the $N-1$ angles $\theta_j$ will span the new search space, of one
dimension less.

\item A \emph{constrained Fourier series}. If the zero-th frequency is
  left out, a Fourier series, Eq.~\ref{eq:fourier_basis},
  automatically fulfills condition
  (\ref{eq:constraint-zero}). Condition (\ref{eq:zerofield}) is met
  if:
  \begin{equation}
    \tilde{\epsilon}_1 = -\sum_{n=2}^{N/2} \tilde{\epsilon}_n.
  \end{equation}
  To fulfill this condition, a first parameter transformation can be defined by
\begin{eqnarray}
\tilde{\epsilon}_1 \ & =: & -\sum_{n=1}^{N/2-1} \alpha_n \ , \nonumber \\
\tilde{\epsilon}_{(n+1)} & =: & \ \ \alpha_n \ ,\qquad\qquad n=1,\dots ,(N-1)\,.
\label{eq:fourier_trans_1}
\end{eqnarray}
In terms of the new coordinates, it is trivial to see that the fluence is
given by a bilinear expression:
\begin{equation}
\mathcal{F}[\alpha] = \vecs{\alpha}^T S \vecs{\alpha}\,,
\end{equation}
for a $(N-1)\times(N-1)$ symmetric matrix $S$. It can be diagonalized by
performing a new change of coordinates based on an orthonormal matrix $U$:
\begin{equation}
U^TSU =
\left( \begin{array}{ccc}
s_1 &  & 0 \\
 & \ddots &  \\
0 &  & s_{(N-1)}
\end{array} \right)\,,
\end{equation}
and if we now define a final change of coordinates in the form:
\begin{equation}
\beta = LU^T\alpha\,,
\end{equation}
where:
\begin{equation}
L :=
\left( \begin{array}{ccc}
\sqrt{s_1} &  & 0 \\
 & \ddots &  \\
0 &  & \sqrt{s_{(N-1)}}
\end{array} \right)\,,
\end{equation}
then the fluence has the simple form:
\begin{equation}
\mathcal{F}[\beta] = \sum_{n=1}^{N-1} \beta_j^2\,.
\end{equation}
Once we have this form, in order to fix the fluence to a predefined
value $\mathcal{F}_0$ one can once again make a coordinate
transformation to hyperspherical coordinates, and use the $N-2$ angles
as search space.

\end{itemize}

\subsubsection{Optimization algorithms.} 
\label{sec:methodology-numerical}

There are two broad families of optimization algorithms: gradient-free
and gradient-based schemes. We will utilize both in the application
presented below.

\paragraph{Gradient-free.} 
In experimental control experiments, the gradient of the merit
function is seldom available, and the most used gradient-free
algorithms belong to the ``evolutionary'' or ``genetic''
families. These are specifically designed for search spaces with large
number of dimensions, typically discrete.\cite{goldberg-evolutionary,
  schwefel-evolutionary}

However, in our code we have opted for two different schemes, which
are sufficient for a moderate number of continuous degrees of freedom:
the classic simplex algorithm of Nelder and Mead,\cite{nelder-mead}
and Powell's NEWUOA algorithm,\cite{newuoa} newer and more efficient.

\paragraph{Gradient-based.} 
If the control function is described in any parameterised
representation, then we have used a standard conjugate gradient
algorithm, the Broyden-Fletcher-Goldfarb-Shanno
variant.\cite{fletcher-cg}

However, if the control function is represented directly in real time
(which usually implies a large number of degrees of freedom), a number
of different algorithms that were specifically developed within the
field of QOCT (or adapted to it) have been proposed. These can provide
very fast convergence, if they are applicable. In particular, very
succesful techniques are the Krotov method\cite{somloi199385} and the
monotonically convergente techniques proposed by Zhu and
collaborators.\cite{zhu19981953, zhu1998385} In some of the examples
given below, we will use one of these latter
techniques.\cite{zhu19981953}

\section{Results: Control targets based on the forces}
\label{section:forces}

During the breaking of a bond, the forces that act on the two
separating nuclei should have more or less opposite directions in
space, i.e., a naive but reasonable attempt to define a bond-breaking
target is to do it in terms of the forces: one can attempt the
maximization of the force difference between the nuclei that must be
separated, and the minimization of the forces between the nuclei
remaining in each fragment. In this section we describe a first
attempt in which the target includes the value of the forces only at
the end of the action of the pulse -- it is, therefore, a static
target.
\begin{figure}
  \centering
  \includegraphics[width=0.60\columnwidth]{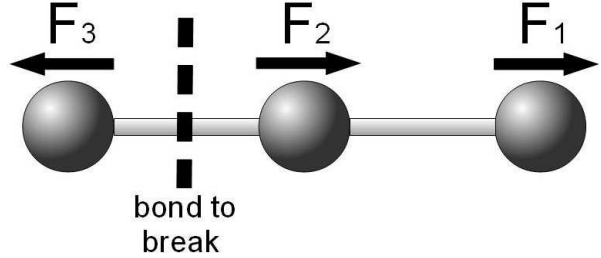}
  \caption[Force control target: Sketch of the 1D model]
   {Sketch of the 1D test model. The direction of the arrows indicates the direction of the force optimization.}
  \label{fig:force_model}
\end{figure}

We will assume that the laser pulse is short, so that during its
action the nuclei do not move significantly; therefore the
optimization calculations will be performed with frozen nuclei. The
idea is that the pulse should be able to place the electrons on a
dissociating state. Later, the optimized pulse will be tested without
the fixed nuclei restriction. In this ``bond breaking test run'',
therefore, the calculation was based on the mixed quantum-classical
description described earlier.

We used a simple 1D model of a triatomic Hydrogen molecule (see
Fig.~\ref{fig:force_model}), with two non-interacting electrons. The
electron-nucleus interaction is modelled with a soft Coulomb
potential:
\begin{equation}
\label{eq:soft_pot}
v_{\rm nuc}(x,x_i) = -\frac{1}{\sqrt{(x-x_i)^2+1}}\,, 
\end{equation}
where $x$ is the electronic coordinate, and $x_i$ is the nuclear
position of nucleus $i$. Since we have two independent electrons
evolving in a spin-independent Hamiltonian, we can assume the system
to be permanently in a singlet state: both the two electrons occupy
the same orbital $\Psi$, which is initially the ground state. It
evolves governed by the Hamiltonian:
\begin{equation}
\hat{H}[\epsilon,t] = -\frac{1}{2}\partial_x^2 + \hat{x}\epsilon(t) +
\sum_{i=1}^3 v_{\rm nuc}(\hat{x},x_i) .
\end{equation}
The temporal dependece of the laser field is determined by the
function $\epsilon(t)$, for which we will consider in this case a real
time representation.

The target functional $F$ will be divided into the object
that truly needs to be optimized, $J_1$, and a penalty function $J_2$:
\begin{equation}
 \label{eq:force_target}
F[\Psi,\epsilon] = J_1[\Psi] + J_2[\epsilon]\,.
\end{equation}
The task of $J_2$ is to prevent unphysically large fluences:
\begin{equation}
J_2[\epsilon] = -\alpha\mathcal{F}[\epsilon] = -\alpha\int_0^T dt \ \epsilon^2(t)\,.
\end{equation}
The constant $\alpha$ is the ``penalty factor''; it is positive, and
it regulates the weight that is put in the low fluence condition.

The definition that we choose for $J_1$ is:
\begin{equation}
\label{eq:force_j1}
J_1[\Psi] = (F_2[\Psi(T)]-F_3[\Psi(T)]) - |F_1[\Psi(T)]-F_2[\Psi(T)]|^2\,,
\end{equation}
where $F_i[\Psi(T)]$ is the force acting on nucleus $i$ at the end of
the pulse action, and is given by:
\begin{eqnarray}\nonumber
F_i[\Psi(T)] &=& - 2 \langle \Psi(T)\vert \partial_{x_i}v_{\rm nuc}(\hat{x},x_i)\vert \Psi(T)\rangle +
\sum_{j\ne i}\frac{Z_iZ_j(x_i-x_j)}{|x_i-x_j|^3}
\\
&=& \int dx \ n(x,T)\partial_{x}v_{\rm nuc}(\hat{x},x_i) + 
\sum_{j\ne i}\frac{Z_iZ_j(x_i-x_j)}{|x_i-x_j|^3}\,.
\end{eqnarray}
This definition of $J_1$ attempts to maximize the force difference
between nucleus 2 and 3, and minimize the force between nucleus 1 and
2. There are some parameters in this expressions that one can
experiment with: the second term in the right hand side of
Eq.~\ref{eq:force_j1} could be multiplied by a weighting factor, or
the square could be eliminated or changed by other exponent. Note that
this type of force target is an explicit functional of the density
$n(x,T) = 2\vert\Psi(x,T)\vert^2$-- this is not so relevant in the
independent electrons approximation taken in this case, but it is in
the Kohn-Sham case that will be discussed later.

We must now adapt the QOCT equations (\ref{eq:schroedinger-1}),
(\ref{eq:schroedinger-2}), (\ref{eq:lambda-1}), (\ref{eq:lambda-2})
and (\ref{eq:qoct-functional-derivative}) to this particular
case. Schr{\"{o}}dinger's equation, together with its initial
condition, (\ref{eq:schroedinger-1}) and (\ref{eq:schroedinger-2}),
obviously do not change. The evolution equation for the auxiliary
$\chi$ wave function is in this case given by:
\begin{eqnarray}
i\frac{\partial \chi[\epsilon]}{\partial t} (x,t)  & = & \hat{H}[\epsilon,t] \chi[\epsilon](x,t) \,,
\\
\chi[\epsilon](x,T) & = & O(x)\Psi(x,T)\,,
\end{eqnarray}
where
\begin{eqnarray}
\nonumber
O(x) & =  & \partial_x[v_{\rm nuc}(x,x_2)-v_{\rm nuc}(x,x_3)] 
\\
& &  -2[F_1[\Psi(T)]-F_2[\Psi(T)]] \partial_x[v_{\rm nuc}(x,x_1)-v_{\rm nuc}(x,x_2)].
\end{eqnarray}
Finally, Eq.~(\ref{eq:qoct-functional-derivative}) takes now the form:
\begin{equation}
\frac{\delta G}{\delta\epsilon (t)} =
-2\alpha\epsilon(t)
+2\mathrm{Im} \langle \chi[\epsilon](t) \vert
\hat{x} \vert \Psi[\epsilon](t)\rangle\,.
\end{equation}
At the maxima, this functional derivative is null, and therefore the
solution field will be given by:
\begin{equation}
\epsilon(t) = \frac{1}{\alpha}
\mathrm{Im} \langle \chi[\epsilon](t) \vert
\hat{x} \vert \Psi[\epsilon](t)\rangle\,.
\end{equation}

\begin{figure}
  \centering
    \includegraphics[width=0.49\columnwidth]{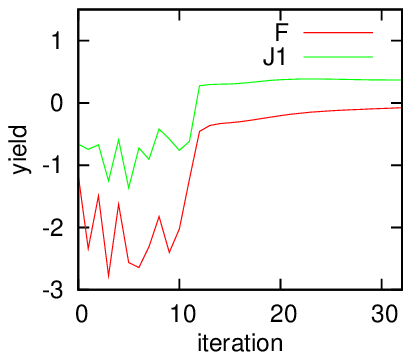} 
    \includegraphics[width=0.49\columnwidth]{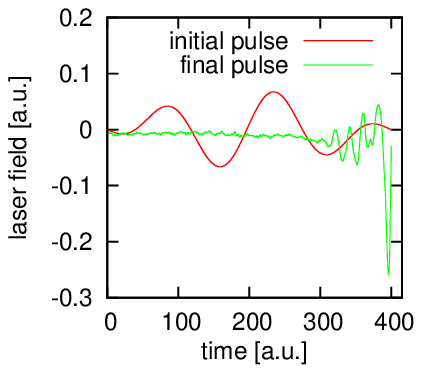} 
  \caption{\emph{Left:} Convergence plot of the optimization run. 
    \emph{Right:} Initial and optimized laser pulses. The optimized pulse
    corresponds to iteration step 30.}
\label{fig:force_conv}
\end{figure}

In order to solve these equations, we chose the algorithm
of Zhu and Rabitz.\cite{zhu1998385} This is a strictly monotonically
convergent algorithm, as long as the target functional has the form
of an expectation value, a condition that does not hold in our case.
The algorithm requires an initial guess, which is then iteratively
improved; we chose a sine wave with sine-shaped envelope (see
Fig.~\ref{fig:force_conv}):
\begin{equation}
\epsilon^{(0)}(t) = A_0 \sin\Big(\frac{\pi}{T}t\Big)\sin(\omega_0 t).
\end{equation}
We used a laser pulse duration of $T=400 {\rm a.u.}$ and an amplitude
of $A_0=7\cdot 10^{-2} {\rm a.u.}$. We tested several frequencies for
the initial field: $\omega_0=(1,2,\dots,9)\cdot 10^{-2} {\rm a.u.}$
(note that the final yield will depend on the choice of the initial
guess).

\begin{figure}
    \includegraphics[width=0.49\columnwidth]{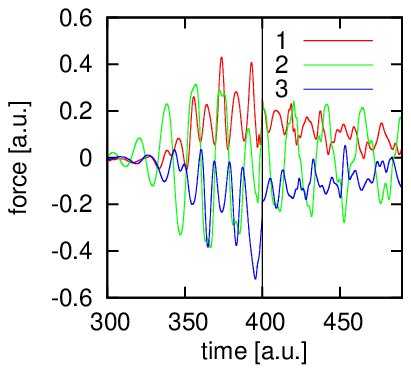}
    \includegraphics[width=0.49\columnwidth]{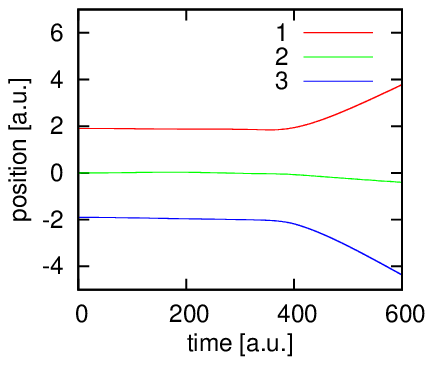}
    \caption{\emph{Left:} Forces on each nucleus during the bond
      breaking test run. The vertical black line indicates the end of
      the laser pulse. \emph{Right:} Position of the nuclei during the
      bond breaking test run.}
\label{fig:force_compare}
\end{figure}

We found that convergence is by no means guaranteed -- only 3 of the 9
optimization runs showed a convergent behaviour.  Furthermore we
observe that the convergence is not monotonic. This is demonstrated in
Fig.~\ref{fig:force_conv} where we show, on the left panel, the
convergence history for the case $\omega_0=4\cdot 10^{-2} \au$ (all
other cases were qualitatively similar).  The right panel shows the
initial and the converged laser pulse.

We use the latter to check whether or not the bond breaks; we let
evolve the system for $1000$ a.u. (i.e., also after the pulse
vanishes) with moving nuclei. Fig.~\ref{fig:force_compare} displays
the forces and positions of the three nuclei during this process. We
first observe that the forces obtained in the optimization run are
not identical to the forces computed during this bond-breaking test
run, since in this case the nuclei have been free to move during the
short laser pulse. However, the differences were small, which validated
(for this particular case) our static nuclei approximation.
A second important observation is that the amplitudes of the force
oscillations before and after the end of the pulse were of the order
of, and even larger than, the optimized forces at the end of the
pulse.

In Fig.~\ref{fig:force_compare} (right), we observe that we got a
complete atomization of the test model in this run -- which is not the
objective. This negative result was typical of all runs: Either the
test model was still bound and the nuclei just oscillated around their
equilibrium positions for $t>400$ a.u., or we got full atomization, as
in the case presented. This latter case was triggered by a strong
electronic ionization.

In view of the strong force oscillations observed, we may conclude
that the main reason for this negative outcome is the time-independent
character of our control target: the forces have a strong oscillatory
character, and controlling them at a single moment in time does not
suffice. This consideration leads naturally to the subject of the next
section: the definition of the control targets in terms of the full
history of the forces -- their integrated values, or in other words, the
momenta.

\section{Results: Control targets based on the momenta.}
\label{section:momenta}

In this section, we explore the option of defining the target
functional in terms of the momenta of the nuclei at the end of the
pulse. For this purpose, we used the same 1D model defined in the previous chapter. 

The momenta are nothing else than the integrated forces:
\begin{equation}
p_i[\Psi] = \int_0^T\!\!\! {\rm d}\tau F_i[\Psi(\tau)]\,,
\end{equation}
and the definition of the target functional $F$ is simply done by
replacing forces by momenta:
\begin{equation}
\label{eq:momentum_target_fixed}
J_1[\Psi] = (p_2[\Psi]-p_3[\Psi]) - |p_1[\Psi]-p_2[\Psi]|^2\,.
\end{equation}
Qualitatively, however, the problem changes, since $p_i[\Psi]$ are
functionals of the full evolution of the system, i.e. we confront a
time-dependent target. The three cases presented below differ in the
manner in which the laser field is defined or restricted, and on the
optimization algorithm.

\subsection{Gradient free optimization algorithm with fixed nuclei} \label{sec:grad_free_fixed}
In this first case, we used a parameterised representation for the
control function (the electric field), in particular the constrained
sine series (see Section \ref{section:laser-field}): the search space
is spanned by a set of hyperspherical angles $\theta
= \lbrace \theta_j \rbrace$, and therefore the fluence is constant
(making unnecessary the introduction of a penalty function $J_2$).

We test now a gradient-free procedure for the maximization of the
function \mbox{$G[\theta] = F[\Psi[\theta],\theta] =
J_1[\Psi[\theta]]$}, in particular the ``downhill simplex'' method
from Nelder and Mead.\cite{nelder-mead} Each function evaluation
amounts to one forward propagation (the backwards propagations are in
this case unnecessary). As in the previous section, we used very short
pulses and assumed the fixed-nuclei approximation during the pulse
action. The optimization runs were followed by the corresponding
``bond-breaking test runs'', in which the nuclei are allowed to move
to check that the molecule breaks in the intended way.

As an initial guess for the pulse, we used, once again:
\begin{equation}
 \label{eq:e_ini}
\epsilon^{(0)}(t) = A_0 \sin\Big(\frac{\pi}{T}t\Big)\sin(\omega_0 t)\,.
\end{equation}
We performed several calculations with varying values of $\omega_0$:
$\omega_0=(4\dots 19)\cdot 10^{-2}\au$. The amplitude $A_0$ is
adjusted so that all optimizations are performed with the same
(constant) fluence. The propagating time was chosen to be
$T=200\au$. All functions were then expanded in a sine Fourier series,
with frequecies $\omega_n = \frac{\pi}{T}n$ for $n=1,2,\dots 12$. This
means 11 degrees of freedom for the search space, once the
transformation to hyperspherical coordinates was done.

\begin{figure}
  \centering
    \includegraphics[width=0.49\columnwidth]{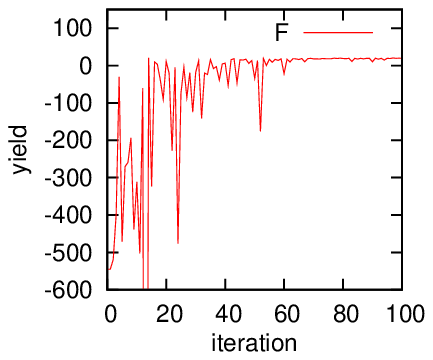} 
    \includegraphics[width=0.49\columnwidth]{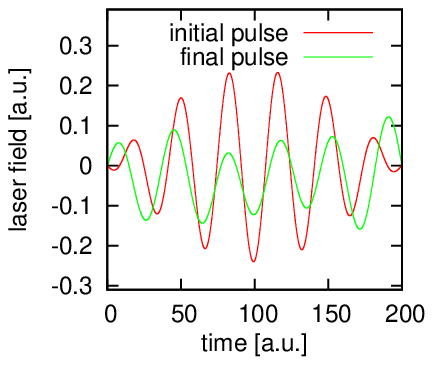} 
  \caption{\emph{Left:} Convergence plot of the
    gradient-free optimization. \emph{Right} Initial and optimized laser pulses.}
\label{fig:momentum_non_conv}
\end{figure}
\begin{figure}
  \centering
  \makebox[0cm]{
    \includegraphics[width=0.49\columnwidth]{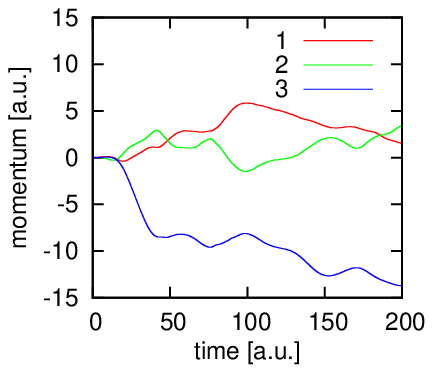} 
    \includegraphics[width=0.49\columnwidth]{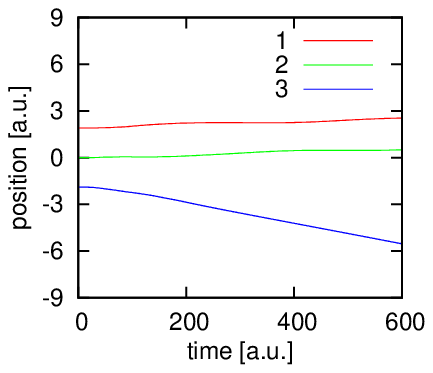}
  }
  \caption{\emph{Left:} Momentum of
    each nucleus during the bond breaking test run. \emph{Right:} Positions of the nuclei during the bond breaking test run.}
\label{fig:momentum_non_momenta}
\end{figure}

All optimizations converged, and of those, half of them led to the
seeked bond destruction. We display results for one of the runs
($\omega_0=19\cdot 10^{-2}\au$), since all of them were qualitatively
similar. Figure \ref{fig:momentum_non_conv} (left) shows the
convergence history of $F$. The right side shows the initial and the
optimized laser pulse (iteration step 100). It is clearly visible that
this optimized pulse does not contain very high frequency components,
compared to the pulse obtained in the forces-based optimization. This
is due to the natural frequency cut-off imposed by the
parameterisation.

The left plot of figure \ref{fig:momentum_non_momenta} displays the
momenta of the nuclei during the bond breaking test run. It is
noteworthy that the momenta did not significantly oscillate for
$t<200\au$, as observed for the forces.  The right plot, in turn,
shows the coordinates of the nuclei during the bond breaking test
run. It is clear that the intended goal was achieved: nucleus 3
dissociates from nucleus 1 and 2, that stay bound. 

Despite the successes, the nuclear movement was not completely
negligible during the action of the laser pulse. This can already be
seen in the right panel of Fig.~\ref{fig:momentum_non_momenta}.  In
order to further study the influence of the nuclear movement, we
performed runs with different pulse durations ($T = 100\au$ and $T =
400\au$). The result is that for $100\au$ many runs succeeded, while
for $400\au$ no run did. We may conclude that (1) constructing the
control target functional in terms of the momenta of the nuclei is an
appropriate approach to the problem of selective bond cleavage, but
(2) the movement of the nuclei is, in general, not negligible when
performing the optimization, unless the laser pulses are very short.

\subsection{Gradient free optimization algorithm with moving nuclei}
\label{sec:grad_free_moving}

The natural next step is therefore to include the ionic motion
during the optimization runs, in order to allow for larger pulse
durations. We have attempted this using exactly the same model and
target definition as in the previous section. The only difference is
that, during the action of the pulse, the dynamic variables include
not only the electronic orbital, but also the nuclear coordinates and
momenta.

The laser pulse was represented in the same way as in the previous
section: the set of hyperspherical angles that describe the fixed-norm
(i.e. fixed fluence) coefficients of a sine series expansion. We
tested, for these runs, in addition to the previously used downhill
simplex scheme, a new gradient-free optimization algorithm: the
NEWUOA\cite{newuoa} scheme. It is based on the construction of a higher
order polynomial approximation to the function that needs to be
optimized.

We used a total pulse length of $T=400\au$, larger than in the
previous case, in order to make the nuclear movement clearly
non-negligible. The sine series expansion contained in this case 14
components, making the parameter space of 13 degrees of
freedom. Several initial guesses of the form (\ref{eq:e_ini})
were tried, with $\omega_0=(3\dots 9)\cdot 10^{-2}\au$.  Each initial
pulse was then optimized with the two maximization algorithms. We
observed a much faster convergence (roughly double) with the NEWUOA
algorithm. All tests, no matter what maximization algorithm was used,
were successful: the optimized pulse led to the breaking of the
selected bond. We describe the results obtained for the case
$\omega_0=6\cdot 10^{-2}\au$ (since all other cases showed a similar
behaviour).

\begin{figure}
  \centering
  \makebox[0cm]{
    \includegraphics[width=0.49\columnwidth]{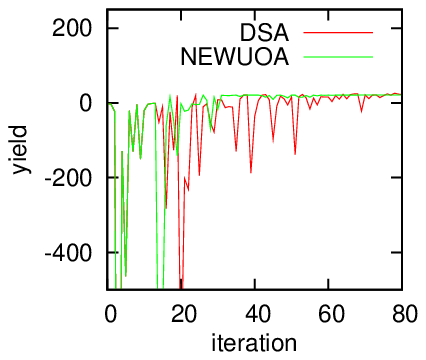}
    \includegraphics[width=0.49\columnwidth]{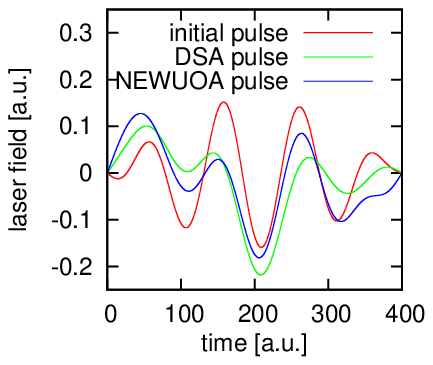}
  }
  \caption{\emph{Left:} Convergence plot for the two different optimization algorithms. DSA stands for downhill simplex algorithm.
           \emph{Right:} Initial and final laser pulse, for both the NEWUOA and the downhill simplex algorithms.}
\label{fig:momentum2_convergence}
\end{figure}

The left plot of Fig.~\ref{fig:momentum2_convergence} compares both
optimization algorithms. Clearly, the NEWUOA algorithm finds the
maximum much faster. The right plot compares the optimized laser
pulses. Here we see that the two algorithms found different local maxima -- even if
both achieved the attempted goal: the breaking of the
selected bond. 

\begin{figure}
  \centering
    \includegraphics[width=0.49\columnwidth]{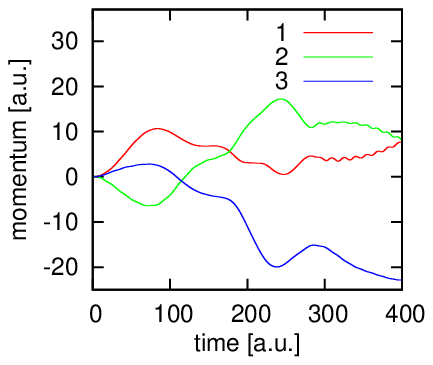}
    \includegraphics[width=0.49\columnwidth]{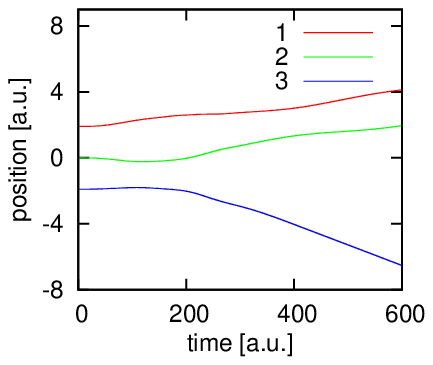}
  \caption{\emph{Left:} Momentum of each nucleus during the bond breaking test. 
           \emph{Right:} Nuclear coordinates, for the same calculation.}
\label{fig:momentum2_momentum_conv}
\end{figure}


The left plot of Fig.~\ref{fig:momentum2_momentum_conv} displays the
momenta of the nuclei during the bond breaking test run (performed
with the laser pulse of iteration step 40 of the NEWUOA optimization).
The right plot displays, in turn, the coordinates. The selected bond
is broken quickly. We note that a certain degree of ionization
occurred in all runs (the final charges oscillated between 1.1 and
$2.0\au$) In all cases, most of the charge remained in the dimer
fragment, permitting its stability. We can conclude that the inclusion of the movement of the nuclei in
the optimization runs solves the problems found in the
previous section, when the pulse durations are not very
short.

\subsection{CG optimization for fixed nuclei}
\label{sec:cg_oct}

A target constructed in terms of the momenta can also be handled with
a gradient based algorithm, for which the QOCT equations are
needed. This subsection describes such calculation for the same model
used in the previous two subsections. Note, however, that the QOCT
equations presented above are valid for a quantum system, not for a
mixed quantum-classical one. Therefore, the nuclei must be frozen
during the optimization, and in consequence we are restricted once
again to short pulses ($T=200\au$, a case for which we saw that the
frozen nuclei approximation is justified).

The laser pulse was represented by the Fourier series
(\ref{eq:fourier_basis}), and further constraints (constant fluence,
zero average field) were then implemented as described in
Section~\ref{section:laser-field} -- the parameter set is then a set
of hyperspherical angles $\theta$. We slightly changed the definition
of the target:
\begin{equation}
J_1[\Psi] = (p_2[\Psi]-p_3[\Psi]) - 10|p_1[\Psi]-p_2[\Psi]|\,,
\end{equation}
in order to have linear dependence with respect to the momenta for the
two terms in the right hand side, since we observed that this choice
usually provides better convergence. The factor ``10'' can also be
changed, and regulates the weight that is placed on the minimization
of the momenta difference between those atoms that must remain bound.

Due to the time-dependent nature of the target, Eq.~\ref{eq:lambda-1}
is now inhomogeneous, and the evolution of the auxiliary wave
function $\chi$ is governed by the following equations:
\begin{eqnarray}
i\frac{\partial \chi[\theta]}{\partial t} (x,t)  & = & \hat{H}[\theta,t] \chi[\theta](x,t) 
+ \frac{\delta J_1}{\delta \Psi^*[\theta](x,t)}\,,
\\
\chi[\theta](x,T) & = & 0\,.
\end{eqnarray}
The gradient $\nabla_{\theta}{G}[\theta]$ can be calculated by
Eq.~(\ref{eq:qoct-gradient}).\cite{note-kevin-thesis} This gradient
can then be used to perform a conjugate gradients\cite{fletcher-cg}
optimization.

We performed a number of runs with this scheme, and the results did
not differ qualitatively of the results obtained with the
gradient-free algorithm: partial ionization, and successful
bond-breaking in about half of the runs. The purpose of these
calculations was to make a comparison regarding the computational
efficiency, and therefore we only show results corresponding to one
run that was performed with identical parameters with both
optimization schemes.

\begin{figure}
  \centering
  \makebox[0cm]{
    \includegraphics[width=0.49\columnwidth]{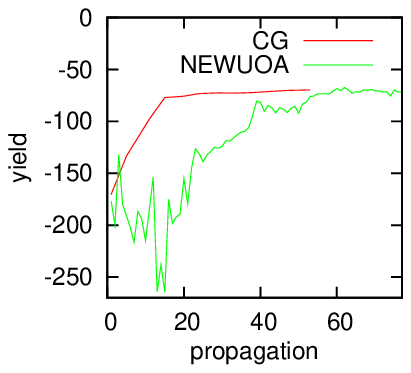}
    \includegraphics[width=0.49\columnwidth]{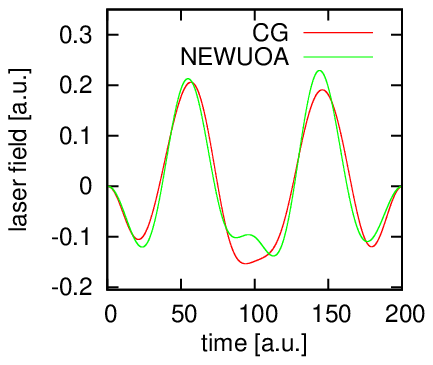}
  }
  \caption{\emph{Left:} Comparison between the CG and NEWUOA optimization. 
           \emph{Right:} Comparison of the optimized laser pulses.}
\label{fig:momentum4_convergence}
\end{figure}
The left plot of Fig.~\ref{fig:momentum4_convergence} compares the
convergence for the two methods. The NEWUOA algorithm reached the
maximum after about 60 propagations, whereas the CG method just needed
about 25 propagations. This was typical, in all runs, the NEWUOA
method needed about twice the computing time to reach convergence.
(note that in the CG case, each propagation corresponds to either a
backwards or a forwards propagation, which require roughly the same
computer time).

The right plot of Fig.~\ref{fig:momentum4_convergence} shows the
optimised laser pulses for both methods. One can see that the two
pulses look rather similar. Nevertheless, there are some small
differences in the optimised pulses which became noticeable in the
ionisation of the system: while the electronic charge decreased to
about $1.65\au$ when irradiating the system with the pulse obtained
with the CG optimisation run, we got a decrease of the electronic
charge to about $1.3\au$ when the NEWUOA pulse was used.

We can conclude that a gradient-based technique such as CG is also
applicable to this problem, and is even more efficient, despite the
complications due to the necessity of backwards propagating an
inhomogeneous Schr{\"{o}}dinger-like equation. Unfortunately, the
scheme cannot yet be applied to longer pulses in which the nuclei
should be allowed to move. In those cases, the nuclear equations of
motion must then be included, as well as the electronic quantum
equation, in the OCT formalism. Work along these lines is in progress.

\section{Selective bond breaking of 1D chains}
\label{section:chains}

\begin{figure}
  \centering
  \includegraphics[width=0.79\columnwidth]{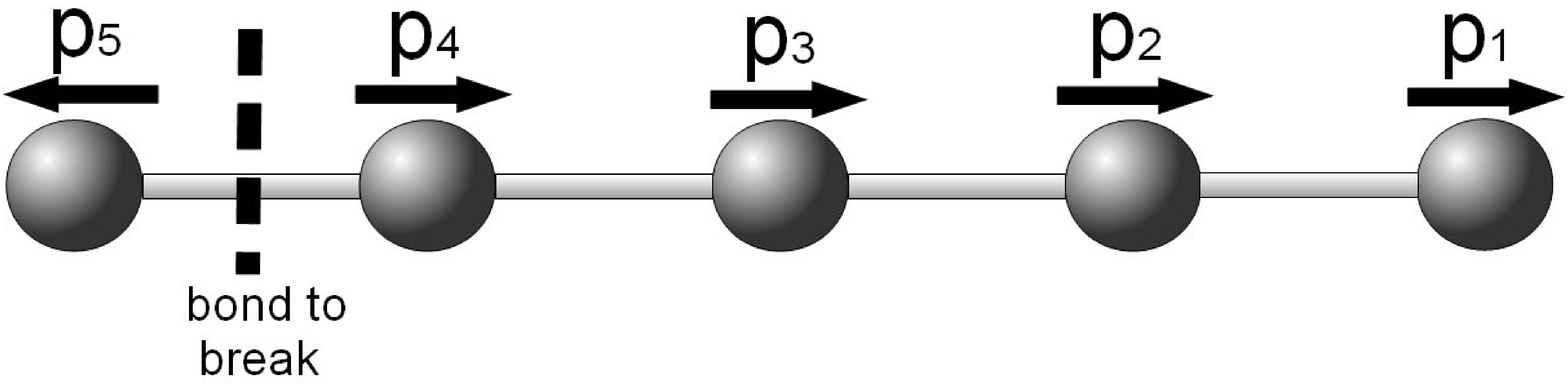}
  \includegraphics[width=0.79\columnwidth]{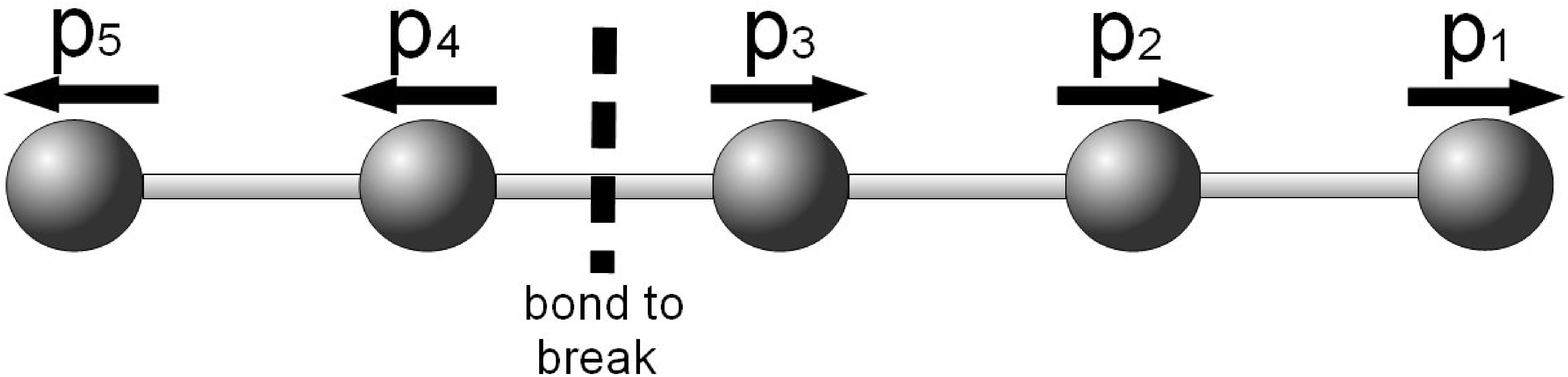}
  \caption{Sketch of the 5-atomic 1D test model.}
  \label{fig:chain1_model}
\end{figure}

A more stringent test on the methodology consists of attempting to
obtain different sized fragments in longer 1D atomic chains. We now
show calculations of five equal mass atom chains, for which we attempt
to break the chain into either 4+1 or 3+2 fragments (see
Fig.~\ref{fig:chain1_model}). The chain consists of 5 Hydrogen nuclei;
as in previous section, they interact with the electrons through a
soft Coulomb potential. We place four non-interacting electrons; this
means that instead of one single wave function $\Psi$, we now have two
doubly occupied orbitals $\psi_1,\psi_2$.
The construction of the control target was based on the same ideas 
discussed earlier: maximising or minimising momenta differences. 
For example, for the 4+1 cleavage attempt:
\begin{eqnarray}
\nonumber
J_1[\psi_1,\psi_2] & = & (p_4[\psi_1,\psi_2]-p_5[\psi_1,\psi_2]) 
\\\label{eq:chain1_target}
& & - 10 \sum_{i=1}^3|p_i[\psi_1,\psi_2] - p_{i+1}[\psi_1,\psi_2]|\,,
\end{eqnarray}
whereas for the 3+2 case:
\begin{eqnarray}
\nonumber
J_1[\psi_1,\psi_2] & = & (p_3[\psi_1,\psi_2]-p_4[\psi_1,\psi_2]) 
\\\label{eq:chain1_target2}
& & - 10 \sum_{i=1,i\ne3}^4|p_i[\psi_1,\psi_2] - p_{i+1}[\psi_1,\psi_2]|\,,
\end{eqnarray}

In this case, we considered moving nuclei and we applied a gradient
free optimization by making use of the NEWUOA algorithm. Again, we
used (\ref{eq:e_ini}) as initial pulse with a pulse duration of
$T=400\au$. The pulse was represented by the constrained sine series
and in this case we restricted the parameter search space to 11
hyperspherical angles.  The following initial parameters have been
tested: $\omega_0=(3\dots 6)\cdot 10^{-2}\au$.

\begin{figure}
  \centering
  \includegraphics[width=0.49\columnwidth]{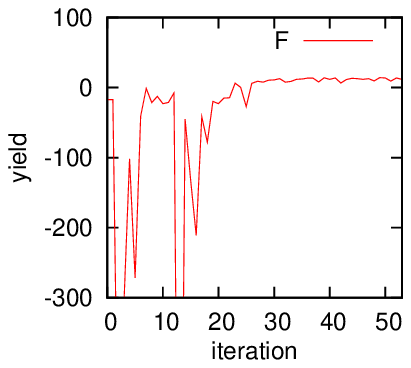}
  \includegraphics[width=0.49\columnwidth]{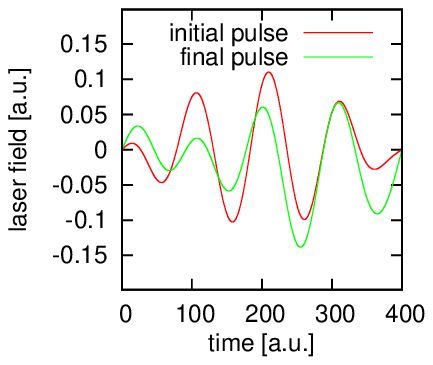}
  \caption{\emph{Left:} Convergence plot (NEWUOA
    algorithm) for the 4+1 chain bond breaking attempt.
    \emph{Right:} Initial laser pulse and optimized
    laser pulse.}
  \label{fig:chain1_convergence}
\end{figure}

\begin{figure}
  \centering
    \includegraphics[width=0.49\columnwidth]{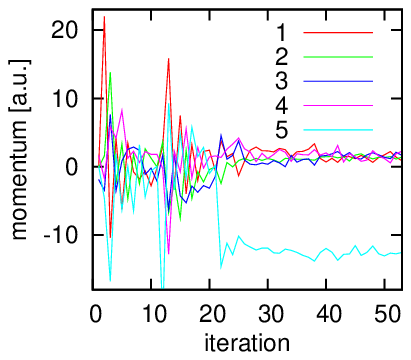} 
    \includegraphics[width=0.49\columnwidth]{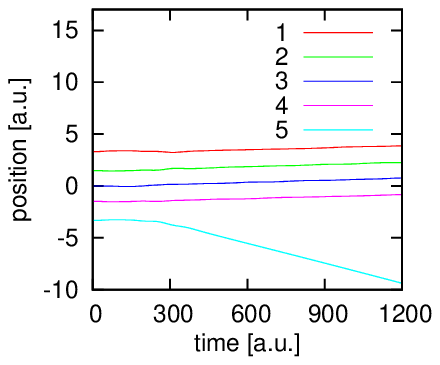}
  \caption{\emph{Left:} Momentum of each nucleus during
    the optimization run, for the 4+1 bond breaking attempt. \emph{Right:} Coordinates of the nuclei during the
    bond breaking test run.}
  \label{fig:chain1_coord}
\end{figure}

For the 4+1 bond breaking attempt, almost all runs were successful
(for two of them the field was too weak to remove any nucleus). There
was no ionization, and all the electronic charge remained by the 4
nuclei, while one proton separated away.  The plots in
Fig.~\ref{fig:chain1_convergence} and
\ref{fig:chain1_coord} correspond to the run with $\omega_0=6\cdot
10^{-2}\au$.  The other successful runs were qualitatively similar.

For the 3+2 bond breaking attempt, the results were different. The
optimization converged for all runs. However, only 2 of 10 bond
breaking test runs were successful. In the other cases, we either
obtained no ionization and unwanted 4+1 separation like in the
previous case, or else substantial ionization and Coulomb explosion of
the full system.

\begin{figure}
    \includegraphics[width=0.49\columnwidth]{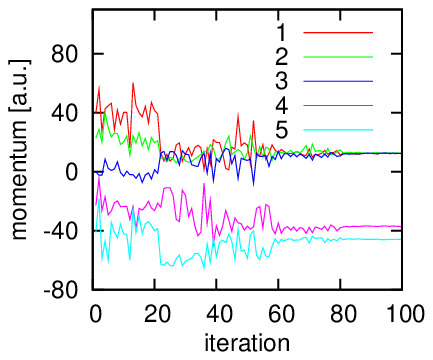} 
    \includegraphics[width=0.49\columnwidth]{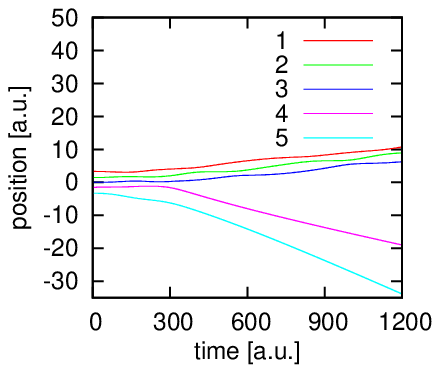}
  \caption{\emph{Left:} Momentum of each nucleus during one of the
    the 3+2 optimization runs. \emph{Right:} Coordinates of the nuclei during the corresponding
    bond breaking test run.}
\label{fig:chain2_coord}
\end{figure}

\begin{figure}
  \centering
  \makebox[0cm]{
    \includegraphics[width=0.49\columnwidth]{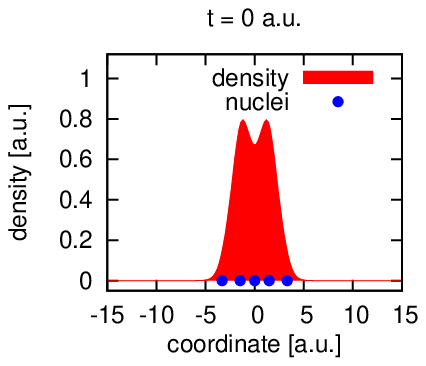}
    \includegraphics[width=0.49\columnwidth]{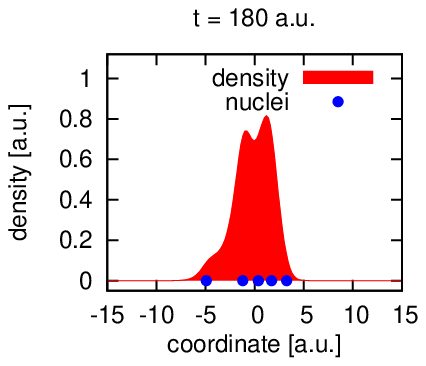} 
    }\\\vspace{0.3cm}
  \makebox[0cm]{
    \includegraphics[width=0.49\columnwidth]{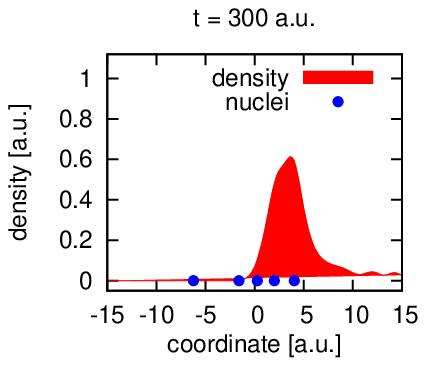} 
    \includegraphics[width=0.49\columnwidth]{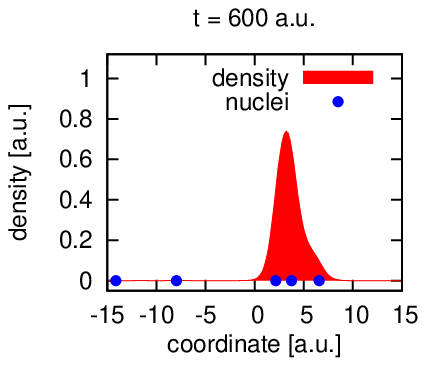}
  }
  \caption{Electronic density, and nuclear coordinates at different times during
    the bond breaking test run for one of the 3+2 chain cleavage attempt.
    The laser pulse duration was $T=400\au$}
\label{fig:chain2_density}
\end{figure}
The plots shown in Fig.~\ref{fig:chain2_coord} correspond to one
successful run, namely that with $\omega_0=6\cdot 10^{-2}\au$.
Fig.~\ref{fig:chain2_density} shows the corresponding electronic
density distribution and the coordinates of the nuclei at different
times. At $t=300\au$, an ionization of the system is observed. In
fact, we found that a certain ionization was needed to remove the two
nuclei. In this particular case, the electronic charge decreased from
$4.0\au$ to $2.3\au$ during the laser pulse.  This ionization
necessarily implied an unwanted effect, namely that nucleus 4 and 5
were not bound to each other anymore after removing them (see plot for
$t=600\au$).

Therefore, for this particular choice of model, search space and
algorithm, the optimization runs did not succeed. However, we expect
that this can be cured in a number of ways, since there is a large
freedom to be explored regarding the definition of the target
functional. For example, the introduction of the ionization in the
definition (prevention or encouragement of ionization) could help to
avoid undesired effects caused by it.  As a final remark, we mention
that we performed further tests with atomic chains with different
masses;\cite{note-kevin-thesis} in those cases, it was found that the
momenta should be substituted by the velocities in order to obtain
better results.

\section{H$_3^+$}
\label{section:h3}

\begin{figure}
  \centering
  \includegraphics[width=0.75\columnwidth]{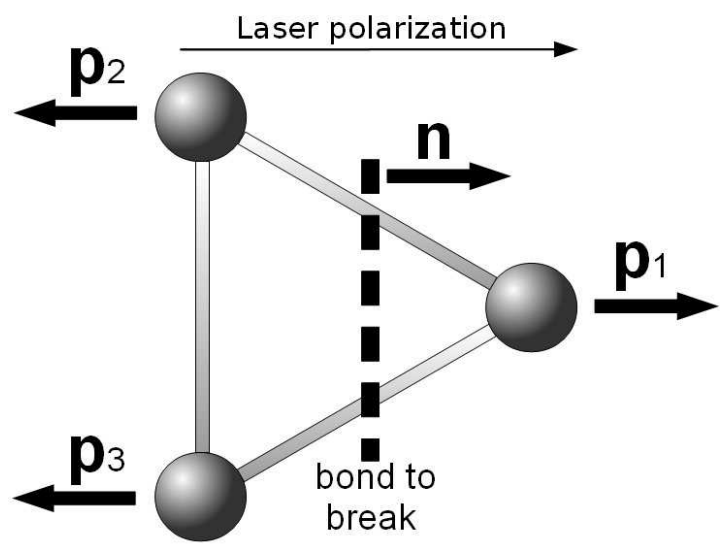}
  \caption[Sketch of \ce{H3^+}]{Sketch of \ce{H3^+}. The dashed line
    indicates the separation plane, where the molecule ought to be
    broken. The normal vector $\vec{n}$ lies in the molecular plane
    and is perpendicular to the separation plane. The laser
    polarization is parallel to $\vec{n}$, and the direction of the
    optimized momenta $\vec{p}_i$ is parallel to $\vec{n}$ as well.}
  \label{fig:h3_sketch}
\end{figure}

The next example is a more realistic molecular description: a 3D
calculation for the \ce{H3^+} molecule, considering interacting
electrons.  Fig.~\ref{fig:h3_sketch} shows the geometry of this
molecule;\cite{h3+} it has an equilateral shape with an edge length of
$1.64\au$.

The two electrons were treated with TDDFT and the exchange-correlation
potential was approximated by the ALDA. The motion of the nuclei was
treated classically. The electron-nucleus interaction was described by
pseudo-potentials -- in this case, obviously, the pseudo-potentials
are not used to remove any core electrons, but as a means to smooth
the Coulomb singularity.

We tried to obtain a laser pulse which removes one particular nucleus,
leaving a bound Hydrogen molecule, by making use of the same kind of
momentum target described above. Fig~\ref{fig:h3_sketch} shows the
directions in which the momenta were optimized; the control target is
defined as:
\begin{equation}
J_1[\Psi] = \vec{n}\cdot(\vec{p}_1[\Psi]-\vec{p}_2[\Psi]) - |\vec{p}_2[\Psi]-\vec{p}_3[\Psi]|\,,
\label{eq:h3_target}
\end{equation}
where $\Psi$ is the Kohn-Sham orbital occupied by the two
electrons. Note that we this functional is an explicit functional of
the density.

We used the gradient free NEWUOA algorithm for the optimization, and
did not neglect the nuclear movement.  The initial pulse was chosen to
be in the form given by Eq.~(\ref{eq:e_ini}), and the pulse duration
was $T=400\au$. In this 3D case, we also have to specify the laser
polarization, which was chosen parallel to $\vec{n}$. The
parametrisation used to represent this laser pulses was the
constrained sine series.

\begin{figure}
  \centering
  \makebox[0cm]{
    \includegraphics[width=0.49\columnwidth]{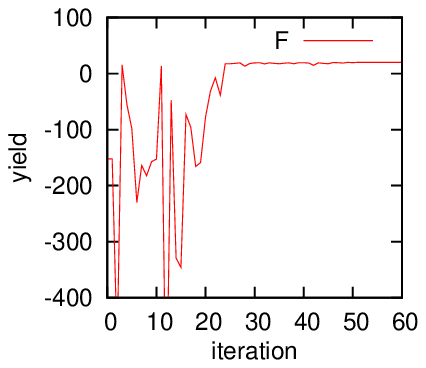}
    \includegraphics[width=0.49\columnwidth]{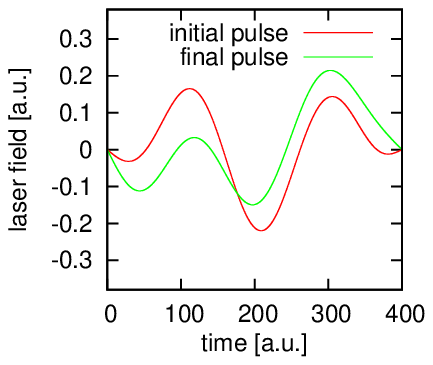}
  }
  \caption{\emph{Left:} Convergence plot for the H$_3^+$ example.
           \emph{Right:} Initial and optimiized (corresponding to iteration step 40) laser pulses.}
  \label{fig:h3_convergence}
\end{figure}

\begin{figure}
  \centering
  \makebox[0cm]{
    \includegraphics[width=0.49\columnwidth]{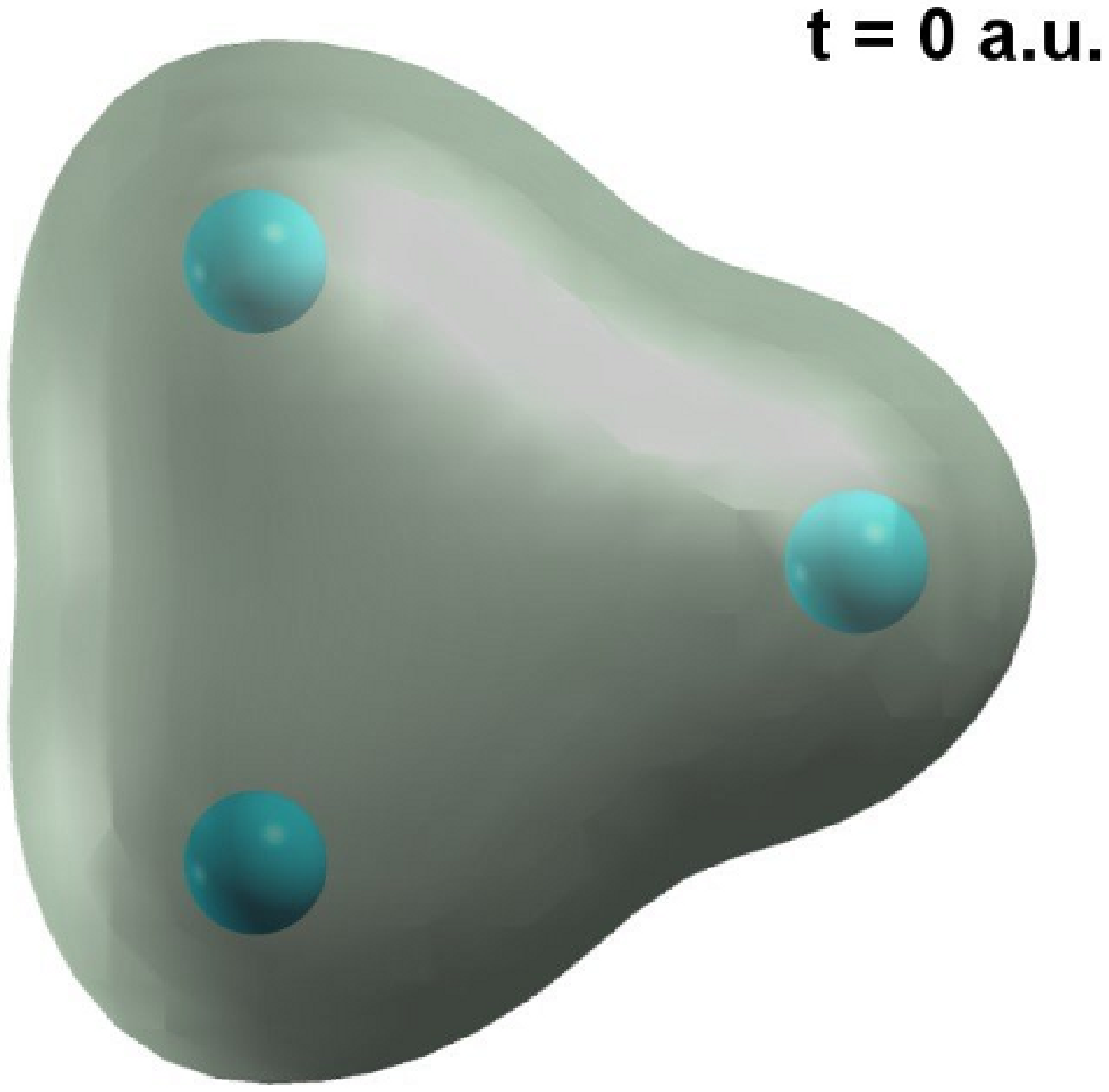}
    \includegraphics[width=0.49\columnwidth]{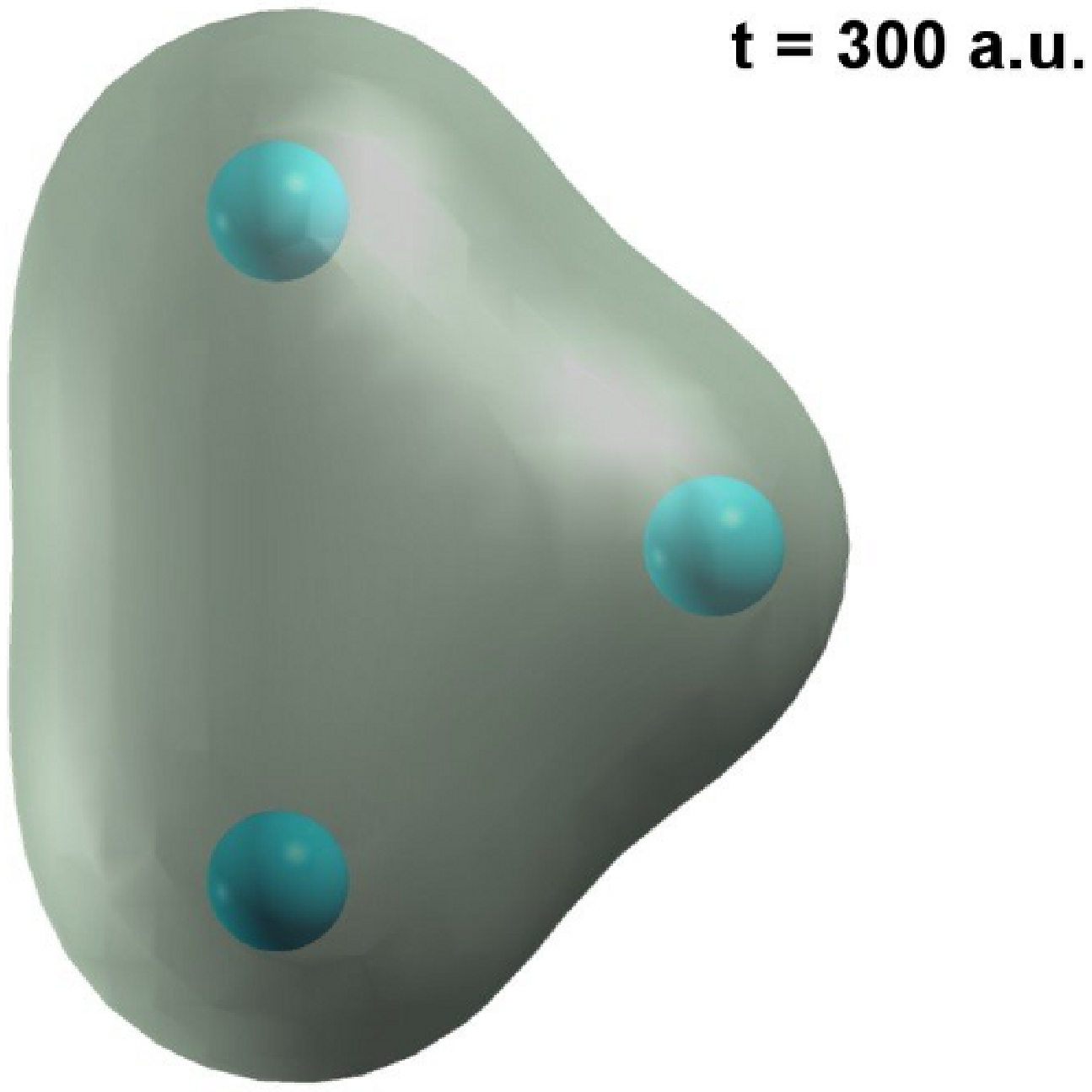}
    }\\\vspace{0.3cm}
  \makebox[0cm]{
    \includegraphics[width=0.49\columnwidth]{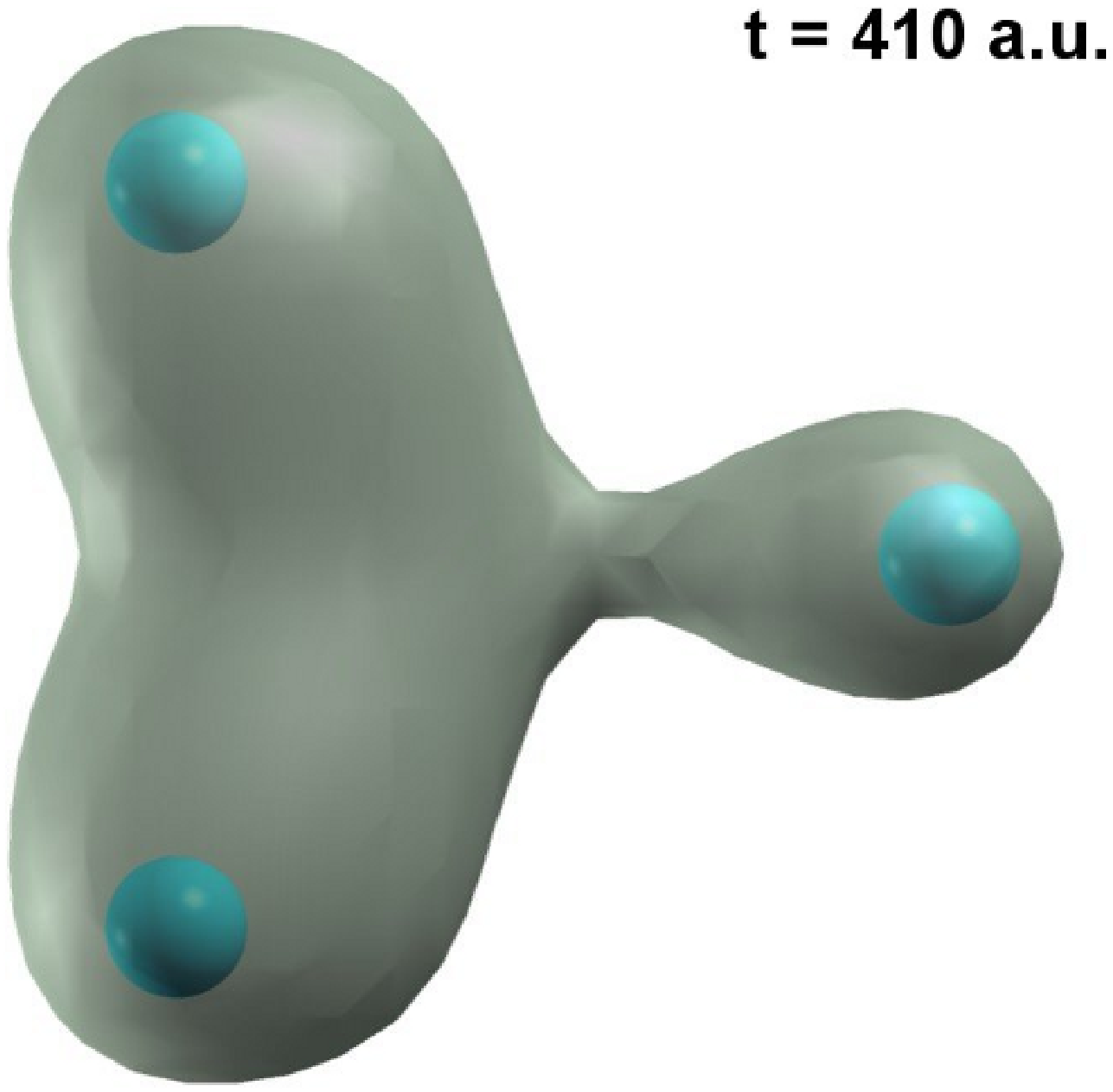}
    \includegraphics[width=0.49\columnwidth]{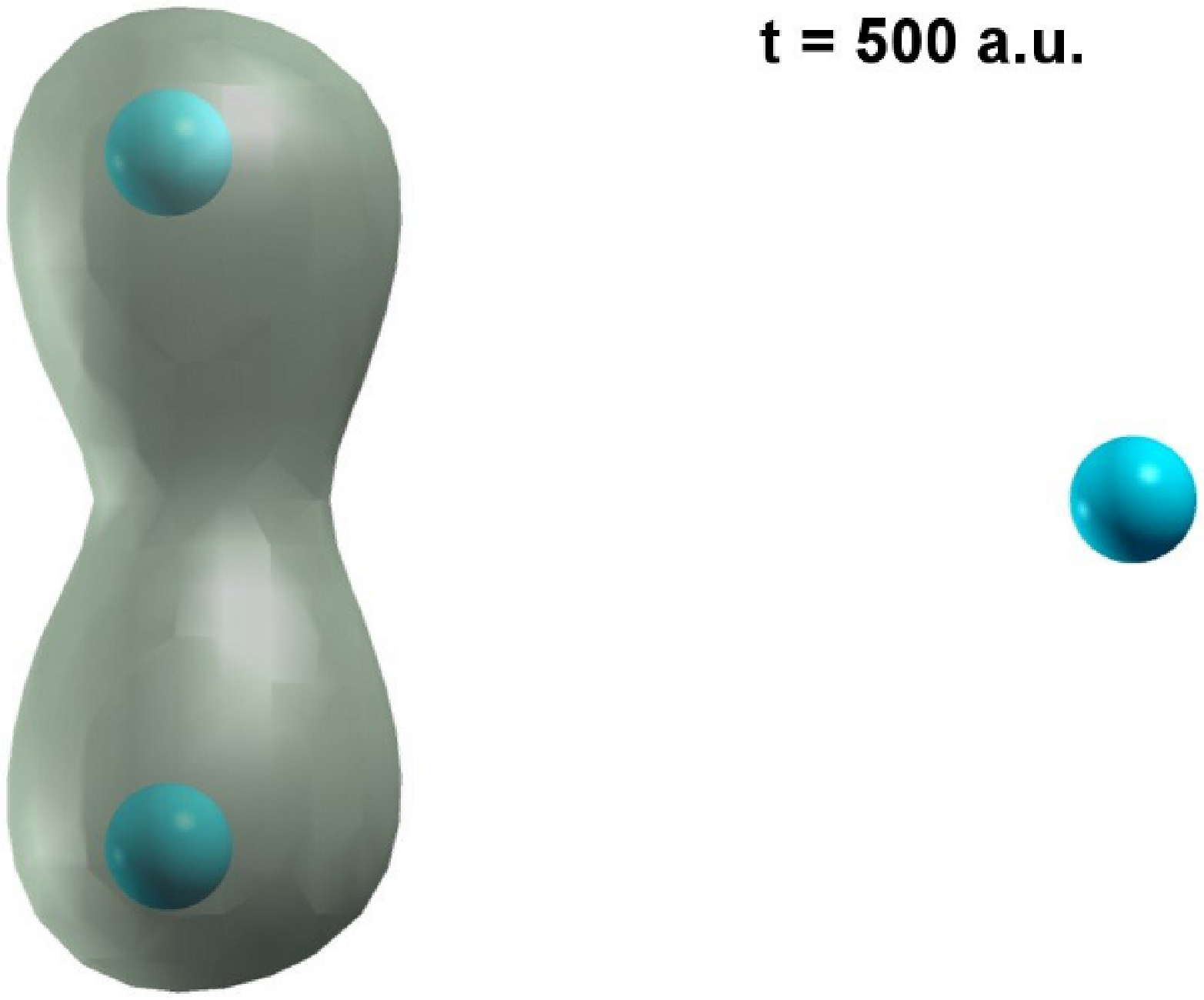}
  }
  \caption{Isosurface plot of the
    electronic density and the corresponding positions of the nuclei
    during the bond breaking test run at different times. The
    isosurface was plotted at a density of $0.07\au$. The laser pulse
    duration was $T=400\au$.}
\label{fig:h3_time}
\end{figure}

We display in Figs.~\ref{fig:h3_convergence} and \ref{fig:h3_time} the
results corresponding to one typical optimization run, corresponding
to an initial guess with $\omega_0=3\cdot 10^{-2}\au$. It can be seen
how the convergence is rather fast, and the obtained pulse cuts the
molecule in the desired way. The electronic charge decreased fromn
$2.0\au$ to around $1.5 \au$

\section{CH$_2$NH$_2^+$}
\label{section:formaldimine}

\begin{figure}
  \centering
  \includegraphics[width=0.9\columnwidth]{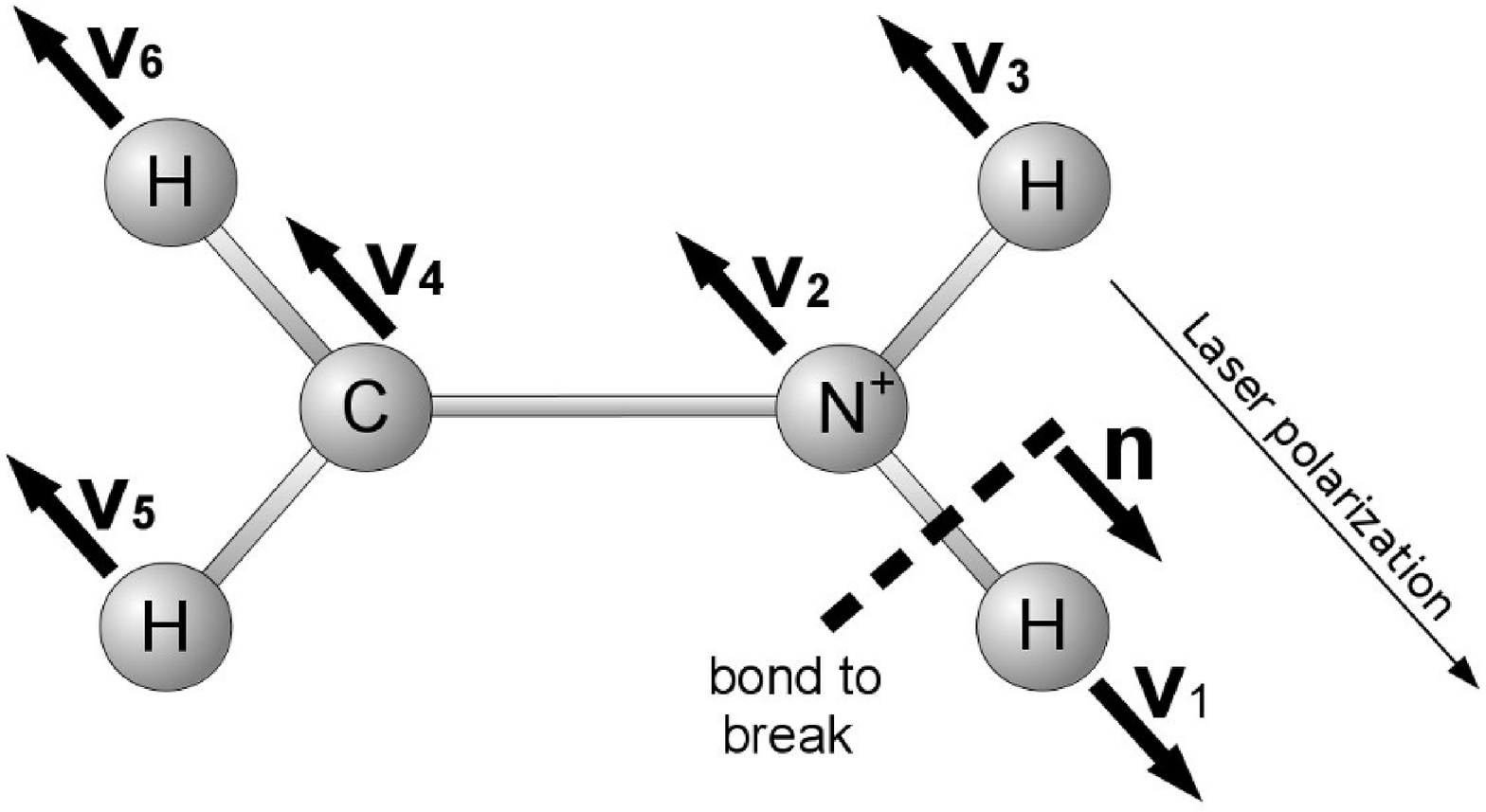}
  \caption[Sketch of \ce{CH2NH2^+}]{Sketch of \ce{CH2NH2^+}. The
    dashed line indicates the separation plane, where the molecule
    ought to be broken. The normal vector $\vec{n}$ lies in the
    molecular plane and is perpendicular to the separation plane. The
    laser polarization as well as the directions in which the
    velocities $\vec{v}_i$ were optimized are parallel to $\vec{n}$.}
  \label{fig:cnh4+_sketch}
\end{figure}

A more complex molecule is \ce{CH2NH2^+}, the ``methaniminium
cation''. The loss of \ce{H^+} as well as \ce{H2} from \ce{CH2NH2^+}
has been extensively investigated, both experimentally and
theoretically.\cite{choi2001114, hvistendahl1985541} Our goal was the
former, the removal of one of the protons, the one that binds to the
Nitrogen nucleus (this process leads to \ce{CH2NH},
``methylenimine'', see Fig.~\ref{fig:cnh4+_sketch}).

We started our calculations from the ground state in
which \ce{CH2NH2^+} has a planar shape (see
Fig.~\ref{fig:cnh4+_sketch}). Then, the simulation of the molecule
dynamics of \ce{CH2NH2^+} was performed similarly to that
of \ce{H3^+}, with the described mixed quantum classical description
on top of TDDFT. The exchange-correlation potential was approximated
by the ALDA. The potentials of the nuclei were described by
pseudo-potentials (in this case, this means that the two core electrons of \ce{C}
and \ce{N} are frozen).

Since we are now working with a molecule that contains nuclei with
different masses, we will define our target in terms of the velocities,
instead of using the momenta (the nuclear labels are defined in Fig.~\ref{fig:cnh4+_sketch}):
\begin{equation}
J_1[n] = \vec{n}\cdot(\vec{v}_1[n]-\vec{v}_2[n]) - 10 \sum_{i=3}^6 |\vec{v}_2[n]-\vec{v}_i[n]|.
\end{equation}
Again, this target functional is an explicit functional of the
electronic density; this is important conceptually since we are using
TDDFT, where the many-body wave function is not easily accessible. In
the previous equation, we have show explicitly this functional
dependence on the density. The normal vector $\vec{n}$ as well as the
laser polarization direction were chosen to be parallel to the bond
axis between the Nitrogen nucleus and the Hydrogen nucleus.

\begin{figure}
  \centering
  \makebox[0cm]{
    \includegraphics[width=0.49\columnwidth]{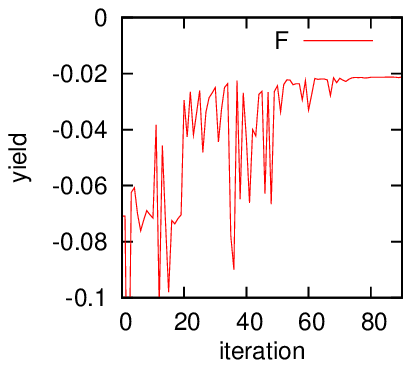}
    \includegraphics[width=0.49\columnwidth]{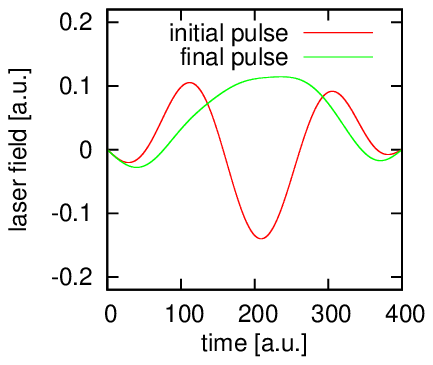}
  }
  \caption{\emph{Left:} Convergence history of the \ce{CH2NH2^+} dissociation attempt, for which the NEWUOA algorithm was used.
    \emph{Right} Initial and optimized laser pulses for this case.}
  \label{fig:cnh4+_convergence}
\end{figure}

\begin{figure}
  \centering
  \makebox[0cm]{
    \includegraphics[width=0.49\columnwidth]{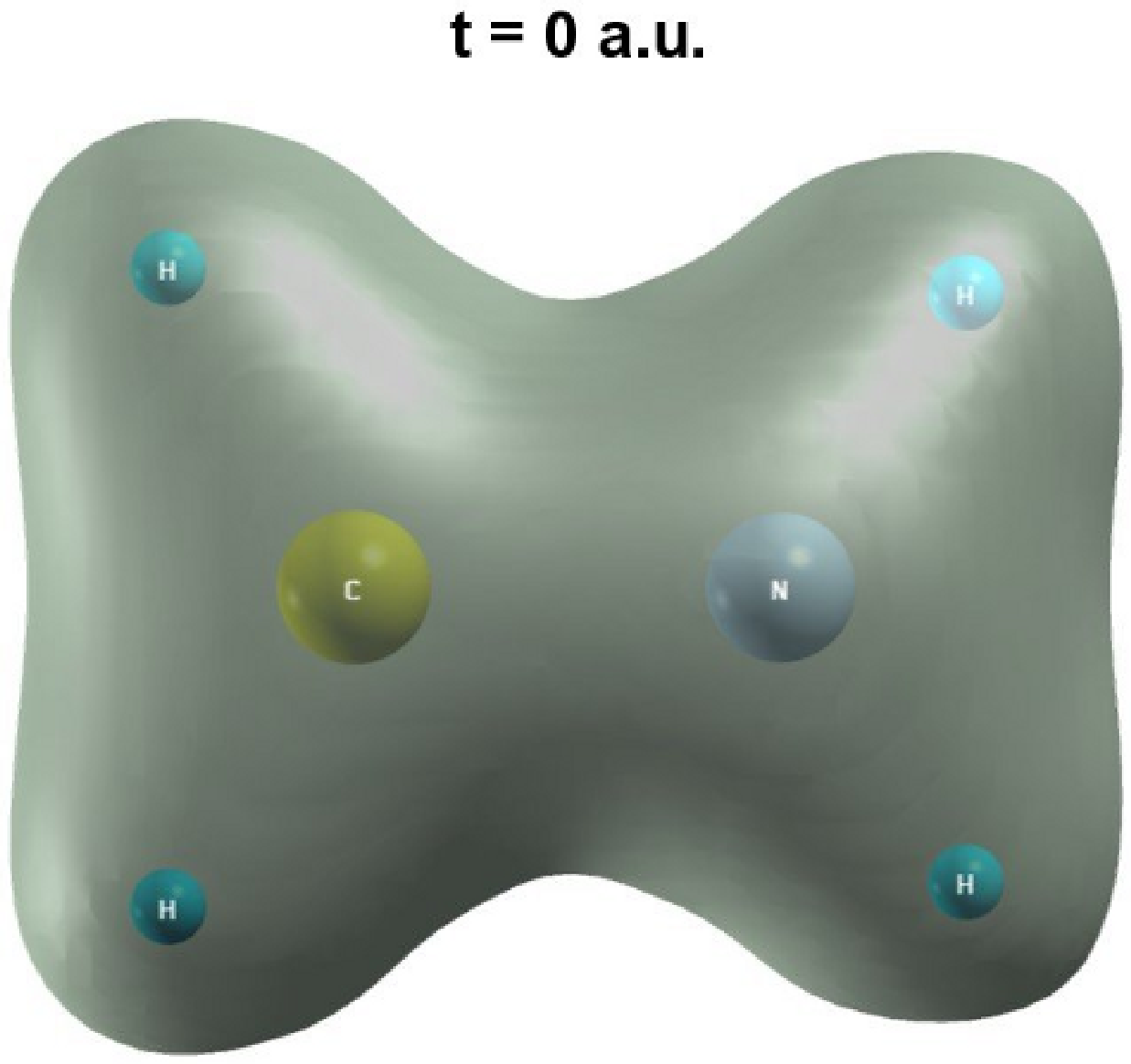}
    \includegraphics[width=0.49\columnwidth]{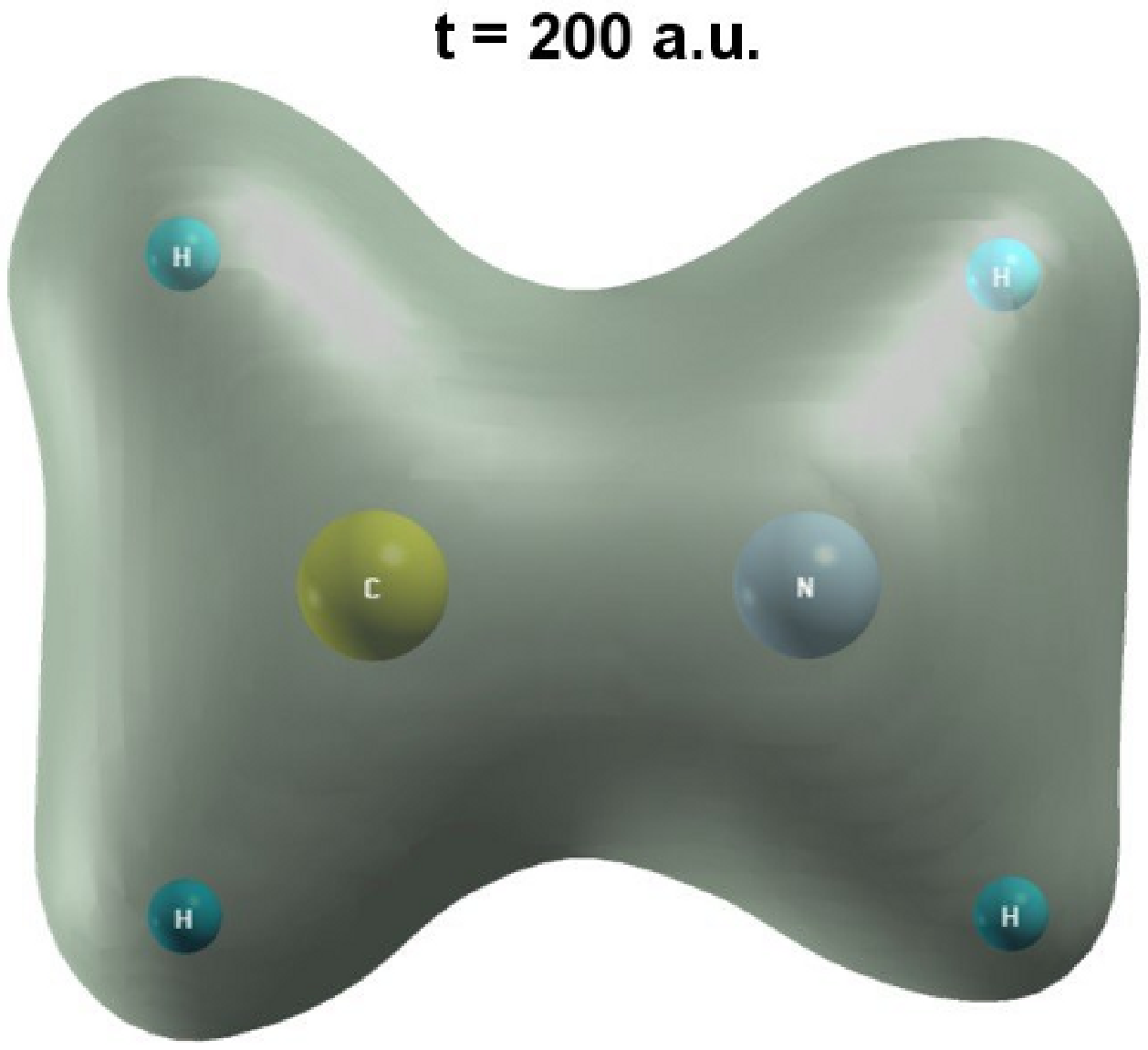}
    }\\\vspace{0.3cm}
  \makebox[0cm]{
    \includegraphics[width=0.49\columnwidth]{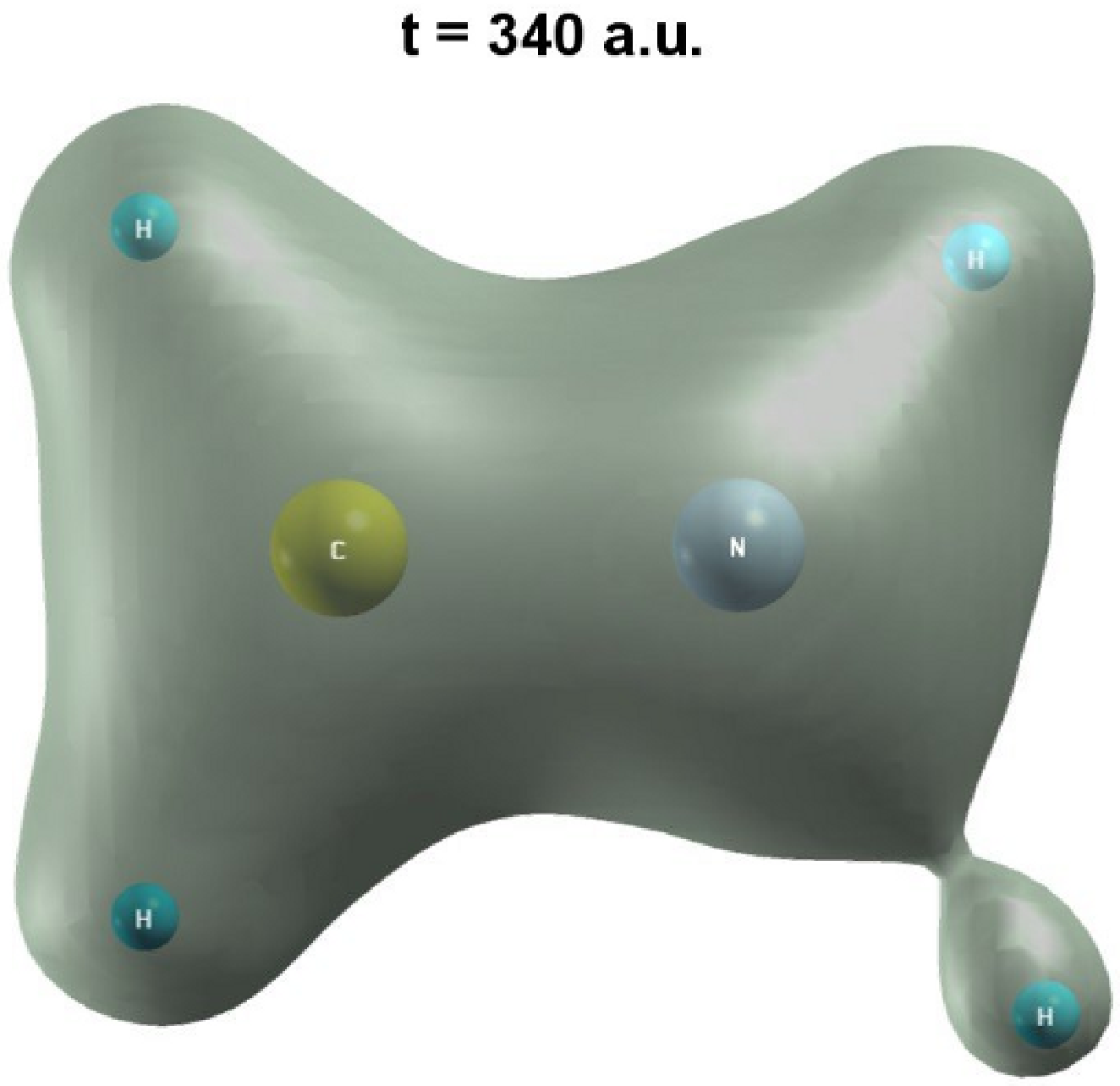}
    \includegraphics[width=0.49\columnwidth]{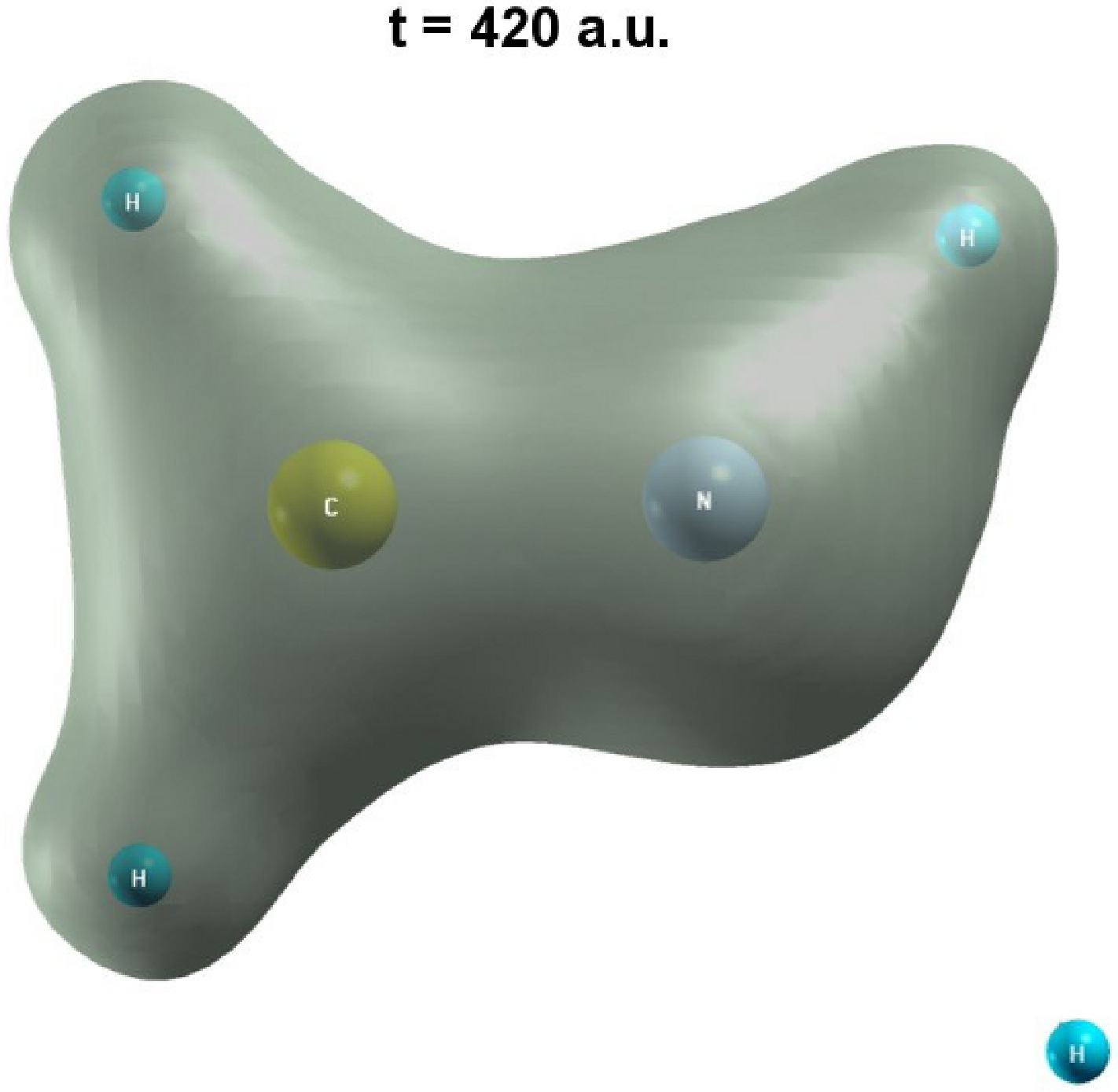}
  }\\\vspace{1.0cm}
  \caption[\ce{CH2NH2^+} run: Electronic density]{Isosurface plot of
    the electronic density and the corresponding positions of the
    nuclei. The isosurface value of the density was $0.045\au$. The
    laser pulse duration was $T=400\au$.}
\label{fig:cnh4+_time}
\end{figure}

Again, we used the NEWUOA algorithm and the form given in
Eq.~(\ref{eq:e_ini}) for the initial pulse; the pulse duration was
$T=400\au$ The electric field was expanded in a sine series, and the
constrained sine series parametrisation was used once again (this
time, with 10 degrees of freedom).  As usual, we performed
optimisations with a number of initial guesses, varying frequencies
and amplitudes (but keeping the fluence constant).  Only one of the
attempts was successful, namely that with the initial frequency
$\omega_0=3\cdot 10^{-2}\au$ The plots in
Fig.~\ref{fig:cnh4+_convergence} and \ref{fig:cnh4+_time} correspond
to this successful run. The electronic charge decreased from $12.0\au$
to $11.0\au$ in this run. In the other cases, the amplitudes of the
optimised electric fields were either too small or too large: too
small electric fields merely led to oscillations of the nuclei around
their equilibrium positions, whereas too large fields, on the other
hand, caused high ionisation, which led to unintended dissociations.

\section{Conclusions}
\label{section:conclusions}

This work addresses the challenge of selective photochemistry by means
of high intensity shaped ultra-short laser pulses. The rapid
experimental advances in forming laser pulses of almost arbitrary
shape call for reliable theoretical tools to predict optimal pulse
shapes for certain predefined tasks. To achieve this goal, our basic
strategy is to combine the mathematical framework of optimal control
theory with a mixed quantum-classical description of the molecular
degrees of freedom: The electronic response of the system is described
from first principles using TDDFT while the nuclear degrees of freedom
are governed by classical equations of motion with Ehrenfest forces
that mediate the coulpling to the electronic degrees of freedom.

The task that the laser pulse is supposed to perform has to be
formulated in terms of a "control target" or "target functional" to be
maximized by the optimal pulse. Usually a given task, like breaking a
selected bond, can be formulated in terms of several possible target
functionals. This is where mathematical intuition and physical
creativity come into play. The mixed quantum-classical description
employed in this work lends itself to formulating the target
functional for bond breaking in terms of the classical nuclear degrees
of freedom. We have explored target functionals based on the classical
forces acting on the nuclei, either considering their value at the end
of the laser pulse, or considering their integrated value over the
full propagation. This latter case means that the target functional
depends on the nuclear momenta at the end of the pulse. The results
show a clear superiority of the momentum-based target functional. This
makes perfect sense because the oscillatory character of the forces
makes their value at a single point in time less relevant than the
integrated values. For molecules with different nuclei, it turns out
to be better to define the targets in terms of the nuclear velocities
rather than the momenta.

After defining the microscopic description of the system and choosing
the control target functional, there is still ample freedom in the
choice of optimization algorithms. We have utilized two fundamentally
different types: gradient-free and gradient based algorithms. The
latter were found, not surprisingly, to perform better. They require,
however, a more elaborate theory, since the gradient (or functional
derivative) calculation involves the backwards propagation of an
auxiliary wave function which is particularly complicated when the
basic equation of motion is non-linear (like in
TDDFT).\cite{arxiv:1009.2241v1}

The calculations presented for H3$^+$ and for CH$_2$NH$_2^+$ clearly
demonstrate that selective bond breaking can be achieved with the
target functionals and optimization algorithms developed in this work.
An immediate task for the future will be the application to larger
molecules. Furthermore, one may consider the definition of refined
target functionals in order to prevent that the removed fragments
break apart later. For example, a term that enhances the electronic
charge localization between the nuclei of the removed fragments could
be included in the target functional. Work along these lines is in
progress.

\section*{Acknowledgements}

This work was partially supported by the Deutsche
Forschungsgemeinschaft within the SFB 658, and by the research project
FIS2009-13364-C02-01 (MICINN, Spain).





\bibliographystyle{elsarticle-num}
\bibliography{rsc}







\end{document}